\PassOptionsToPackage{authoryear,round,sort&compress}{natbib}

\documentclass[aps,prl,reprint,twocolumn,preprintnumbers,longbibliography]{revtex4-2}

\usepackage{graphicx}  
\usepackage{subfigure}
\usepackage{multirow}
\usepackage{placeins}

\usepackage{physics}

\usepackage{fancyhdr}
\usepackage{longtable}
\usepackage{parskip}
\usepackage[T1]{fontenc}
\usepackage{dcolumn}   

\usepackage{bm}        
\usepackage{amsfonts}  
\usepackage{amsmath}   
\usepackage{amssymb}   
\DeclareMathOperator\arctanh{arctanh}
\usepackage{lipsum}

\usepackage{subfigure}	
\usepackage{booktabs}
\usepackage[breaklinks]{hyperref}

\usepackage{multirow}
\usepackage{comment}
\graphicspath{ {figures/} }

\begin{document}
	
	\title{Simulation of Space Platform Charging in Very Low Earth Orbit Using Stochastic Particle Methods}
	
	\author{\textbf{J. Skácel}, P. Parodi, G. Gangemi, F. Bariselli, Z. Bonaventura, T. Magin}
	
	\affiliation {\it von Karman Institute for Fluid Dynamics, Aeronautics and Aerospace Department, Belgium}
	
	\date{July 21, 2025}
	
	\begin{abstract}  
		
		The residual atmospheric drag on satellites in Very Low Earth Orbit (VLEO) has been recognized as a limiting factor for satellite lifetimes. This work focuses on one component of the drag, i.e.,  the ionospheric drag from charged particles, which was observed to influence the overall aerodynamics of satellites. Space platform charging processes are modeled using a stochastic particle method—Particle-in-Cell—with with the possibility to treat the electrons as a Boltzmann fluid, thus reducing the computational time while preserving the accuracy of the physics of charging. Part of the study involves examining non-equilibrium distribution functions, specifically the Kappa distribution, commonly used to describe space plasmas. It has been shown that such distributions can noticeably affect overall space platform charging, thereby enhancing ionospheric drag on satellites using the proposed hybrid models. Nevertheless, it is acknowledged that ionospheric drag might constitute only a fraction of the total drag caused by neutral atmospheric gas.
		
	\end{abstract}
	
	
	\maketitle 

	\section{Introduction}\label{sec:1}
	
	Satellites flying in orbits at low altitudes are exposed to a relative flow of gas from the residual atmosphere that results in a net drag on the satellite. This can essentially determine the lifetime of a satellite due to the decay orbit if there are no means of additional propulsion to balance the resulting drag and stay on a stable orbit. This is especially true for Very Low Earth Orbits (VLEO) or Low Earth Orbits (LEO) when the satellite has a limited fuel; the life expectancy ranges from months to few years depending on the altitude until they decay to thick atmosphere and burn. In recent years, a novel idea of air breathing propulsion systems have been introduced to satellites as a solution to the limited fuel. Nevertheless, in all cases, the study of the drag on the bodies in orbits is required to estimate the needed power for a stable orbit.  
 
	This work aims to explore the physics of the interaction between spacecraft and charged particles in the ionosphere, specifically within VLEO. Employing particle methods, the main objective is to simulate and comprehend the charging processes on spacecraft surfaces, concurrently revealing the resulting electric potential structure around the spacecraft. Electrons, a significant component of the ionosphere, are often approximated as a fluid due to their small mass, known as Boltzmann electron fluid. This project explores and develops such an approximation for numerical simulations, seeking to verify its validity under VLEO conditions through a direct comparison between particle and fluid representations. A main aspect of the project focuses on simulations of a specific phenomenon known as ionospheric drag. This phenomenon is understood as enhanced drag on the charged spacecraft caused by the direct interaction with ionospheric plasma. A greater attention is given to a non-equilibrium distribution function, called Kappa distribution function, that has been utilized in describing the space plasma.

Recent works has shown the importance of ionosphere drag~\cite{lafleur, capon} using numerical computing or experimental teting. It has been shown that ionospheric drag coming from charged particles can become important for given plasma conditions~\cite{lafleur}. Additionally, the application of such drag has been proposed for practical aerobraking~\cite{kleinig}. Stochastic particle methods have been utilized as a main tools to describe the ionospheric drag on solid bodies~\cite{capon2, capon3} while utilizing approximated models for the electrons. Particle methods have been also used to study plasma--body interaction in the solar system~\cite{yaroshenko} and have also been studied for potential hazards on the satellites~\cite{KOONTZ}. Additional works~\cite{marchand, cully} have shown the influence of surface reactions to the space platform charging. A scientific community for space platform charging~\cite{SPINE} has been intensively studying such effects.

In most of the latter works, where fluid approximations were used for electrons, specifically the Boltzmann electron model, the discussion for its validity is missing. The assumption of Boltzmann electrons requires collisions which is usually true for laboratory plasma and its thermal equilibrium state is then described by Maxwell-Boltzmann distribution~\cite{piel,bittencourt}. Space plasma, in ionosphere, can be collisionless compared to typical length scales of satellites and deviations from equilibrium~\cite{ASHIHARA19741201,marif} likely occur. These are described by the suprathermal electrons, having higher energy than electrons described by Maxwell-Boltzmann distribution. In this work, we wish to comment on the validity of such distributions in ionosphere plasma.

The origin of suprathermal electrons in ionosphere are not properly dated but has been reported in outer regions of the Earth's atmosphere~\cite{Livadiotis_2010, Nicolaou_2018} and in solar winds~\cite{stverak} that interact with the ionosphere. These electrons can be essentially described by the Kappa distribution~\cite{Guo_2021,López_2023}. The influence of particles with such a distribution on the space platform charging is not usually discussed, but it is admitted in a work~\cite{sarrailh, Pierrard} to show a difference. We aim to investigate and properly address the influence of Kappa particles on the overall space platform charging and the resulting ionospheric drag using proposed stochastic particle methods.

In summary, this work contributes to the understanding of spacecraft charging in VLEO, striving for accuracy in simulating charging processes and providing insights to refine computational models with practical applications. The project further seeks to address following key questions concerning the challenging conditions of VLEO:
\begin{enumerate}
    \item What is an efficient numerical method to accurately capture both space platform charging and satellite aerodynamics?
    \item What is the effect of thermal non-equilibrium distribution functions of plasma on platform charging?
\end{enumerate}

The work is outlined accordingly. The following~\autoref{sec:2} is devoted to the theory of space charging physics and the relation to the aerodynamics of a satellite -- drag calculation.  In the next~\autoref{sec:3} a particle method is proposed known as Particle-In-Cell (PIC) method for the simulation of plasma--satellite interaction, where we outline the theory and implemented models in the employed software, PANTERA. The~\autoref{sec:4} contains proposed test cases for the verification of the implemented models, 1D plasma sheath in~\autoref{subsec:41} and sphere flight for neutral drag in~\autoref{subsec:42}. Lastly, we present the application in~\autoref{sec:5} with discussion in a case of ionosphere plasma--solid body interaction simulation, using both simple sphere geometry and more complex geometry of a satellite. Lastly, we proceed with conclusion and future work in~\autoref{sec:6}.

\section{Space charging physics}\label{sec:2}
	
\subsection{Ionosphere plasma--solid body interaction in VLEO}\label{subsec:21}
\begin{figure*}[ht]
    \includegraphics[width=0.75\linewidth]{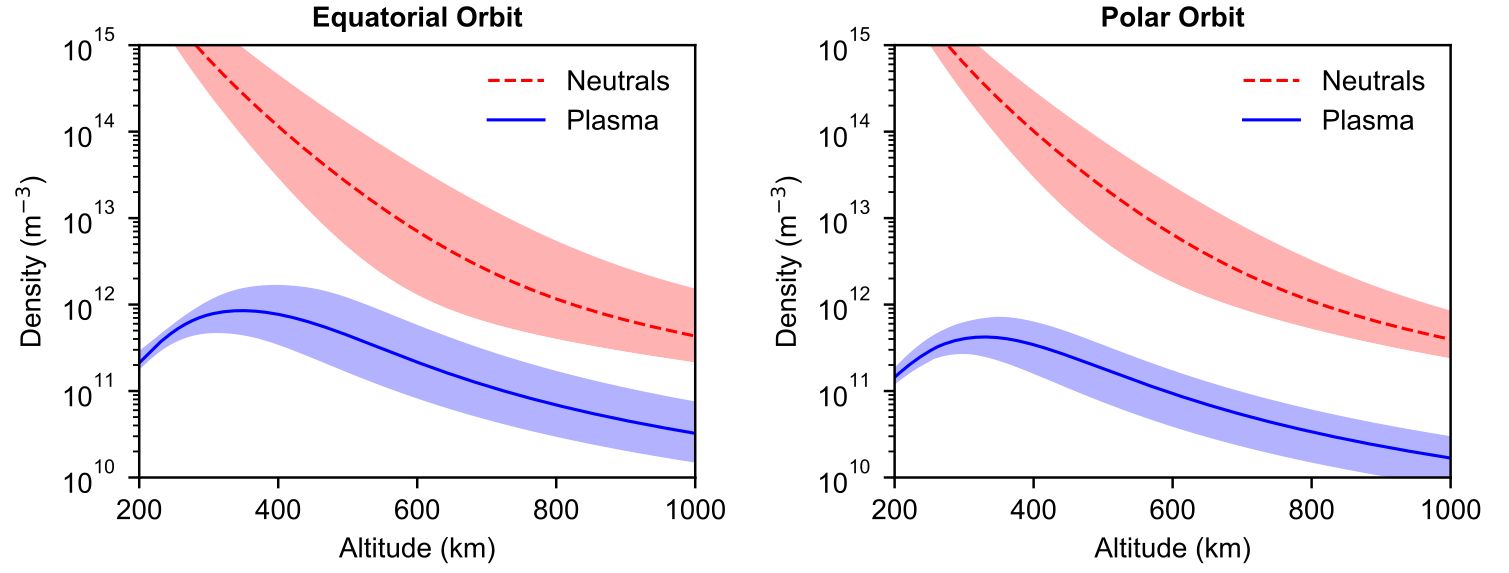} 
    \caption{Neutral and plasma density as a function of altitude. The shown values are averaged through orbits. Picture taken from~\cite{lafleur}.}
    \label{fig:21_density_altitude}
\end{figure*}

Solid bodies flying in a VLEO orbit are exposed not only to neutral particles (atoms and molecules) of the residual atmosphere, but also to a ionized gas that consists of charged particles (ions and electrons) forming the ionosphere, a protective layer that ranges from 100 to 1000 km~\cite{space}. It is also a fact that the peak in the density of charged particles lies in the altitude range from 200 to 300 km~\cite{IRI, lafleur} (example of orbit conditions shown in~\autoref{fig:21_density_altitude}), which belongs to a typical VLEO orbit. 

\begin{figure}[ht]
    \centering
    \includegraphics[width=0.8\columnwidth,trim={0.4cm 0.3cm 0.4cm 0},clip]{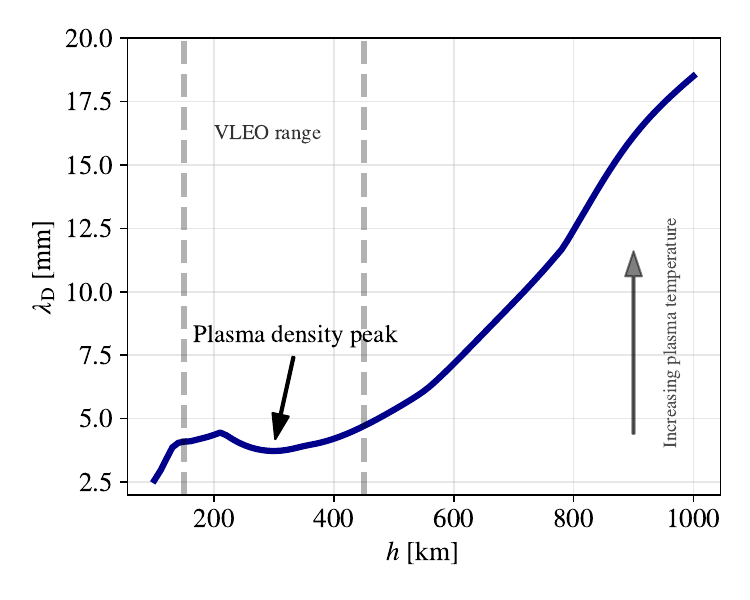}
    \vspace{-4mm}
    \caption{Debye length as a function of altitude for specific orbit (50°LA  50°LO, 01/06/2024 10:00).}
    \label{fig:21_debye_altitude}
\end{figure}

While the physics of neutral praticle's interaction with solid bodies are well understood, the interaction between the ionosphere plasma and a solid body is more complex to describe. The main difference is in the existence of collective behaviour of the charged particles; they interact with each other through electrostatic forces on long distances. However, in a bulk plasma, these forces are effective only over distances of the order of the Debye length $\lambda_{\mathrm{D}}$~\cite{bittencourt}, a characteristic length scale over which any local disturbances in the electric field are screened out by the plasma. It is given by following expression
\begin{equation}\label{eq:21_debye_length}
    \lambda_{\mathrm{D}} = \sqrt{\frac{\varepsilon_0 k T_{\mathrm{e}}}{n_{\mathrm{e}} e^2}}
\end{equation}
where $n_{\mathrm{e}}$ and $T_{\mathrm{e}}$ are the electron density and temperature respectively, $\varepsilon_0$ is the permittivity of vacuum and $k$ is the Boltzmann constant. The screening effect is generally accounted for the electrons due to their low mass, resulting in a quicker response to any disturbances compared to heavy ions. The Debye length is an important plasma parameter that plays a vital role in terms of plasma--solid body interaction~\cite{cully}. A relation of the Debye length with respect to the altitude is presented in~\autoref{fig:21_debye_altitude}. It is shown that the Debye lengths are usually smaller than the typical lengths of satellites in VLEO, while in higher altitudes, the Debye length can be comparable to the dimensions of smallest satellites, e.g. CubeSats.

The electrostatic interaction of charged particles is described by the Poisson equation
\begin{equation}\label{eq:21_poisson_equation}
    - \div(\varepsilon \grad{\phi}) = \sum_{j \in S} q_j n_j
\end{equation}
where $\phi$ is the electric potential, $\varepsilon$ is the permittivity that can generally vary in space. On the right hand side, we have the sum of charge density over all the species $j$ in $S$, while $q_j$ and $n_j$ is the charge and density of species $j$, respectively. The solution for the electric potential $\phi$ is seeked to describe the plasma dynamics in the vicinity of solid bodies. 

Important aspect of the plasma--solid body interaction is the collection of charged particles by the solid body. While different kinds of reaction might occur on the surface (such as surface chemistry, neutralization of ions and photoemission)~\cite{piel,lieberman2005principles}, this works only assumes a simple model of charge attachment to the surface of the solid body; a charged particle collides with the surface and attaches at the collision location. Macroscopically, one can express the amount of particles arriving by the random flux from a particle distribution function. For Maxwell-Boltzmann distributions one can obtain a random flux $J$ in arbitrary direction~\cite{bittencourt}
\begin{equation}\label{eq:21_flux}
    J_j = \frac{1}{4} n_j \underbrace{\sqrt{\frac{8 k T_j}{\pi m_j}}}_{\overline{u}_j}
\end{equation}
where $m_j$ is the mass of species $j$, $T_j$ is the temperature of species $j$, each having having its own energy bath (temperature) and $\overline{u}_j$ is the mean speed. It is clear that in quasi-neutral plasma, the random flux from electrons will be the most dominant due to their low mass ($m_{\mathrm{e}} \lll m_i$, where $m_i$ is the mass of heavy ions). In this case, if a solid body is immersed into a plasma it will initially get charged on negative potential with respect to the free-stream, quasi-neutral plasma. This is followed by further repulsion of electrons from the surface and attraction of positively charged particles towards the body. A steady state is reached when a net charge current balance is reached. In this situation, the solid body is charged on so-called \textit{floating} potential if the body is not conductive~\cite{bittencourt}. While the body can get charged uniformly in quiscent plasma, in orbit conditions, the relative flow of ionosphere plasma introduces non-uniform surface charging that can lead to potential differences across the solid body~\cite{capon2}. Such phenomena is relevant for cases, such as circuit hazards, where the potential difference across surface is sufficient enough to create strong discharges that can damage the satellites~\cite{KOONTZ}, or for the ionosphere aerodynamics of a body, which is the aim of this work for study. 

\begin{figure}[ht]
    \centering
    \includegraphics[width=0.8\columnwidth,trim={0.4cm 0.3cm 0.4cm 0},clip]{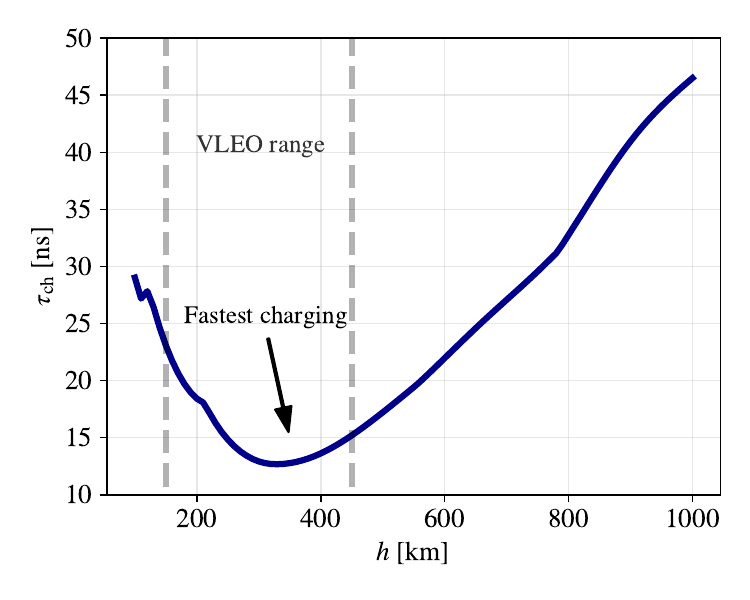}
    \vspace{-4mm}
    \caption{Characteristic charging time as a function of the altitude for specific orbit (50°LA 50°LO, 01/06/2024 10:00).}
    \label{fig:21_charging_altitude}
\end{figure}

A relevant quantity to express the charging process is the characteristic charging time 
\begin{equation}
    \tau_{\mathrm{ch}} = \frac{\lambda_{\mathrm{D}}}{\overline{u}_j}
\end{equation}
This quantity is shown in~\autoref{fig:21_charging_altitude} as a function of the altitude. We can see that the fastest charging processes happens in VLEO orbits.

\subsection{Velocity distributions of charged particles}\label{subsec:22}
Distribution functions belong to the mathematical tools for the statistic description of a plasma state. This function gives us a number of particles within a volume in the phase space~\cite{liboff1969introduction}
\begin{equation}
    \dd N_j = f_j \dd^3\vb{v} \dd^3\vb{r}
\end{equation}
where $N_j$ is the number of particles of species $j$, $f_j$ is the (one-particle) distribution function, $\vb{v}$ is the velocity and $\vb{r}$ is the position vector in the phase space, a 6D space span by the vector $\vb{r}$ and $\vb{v}$. The spatial and temporal characteristics of such a function are generally unknown a priori, with its evolution described by the Boltzmann equation. One can then obtain all the macroscopic quantities, such as density, flux, mean energy, etc. 

\subsubsection{The Maxwell-Boltzmann distribution function}

A state of a plasma in thermodynamic equilibrium can be described by the isotropic Maxwell--Boltzmann (MW) velocity distribution function (VDF) for species $j$ that reads
\begin{equation}\label{eq:22_MW_VDF}
    f_j^{\mathrm{MW}}(\vb{v}) = n_{0,j} \left( \frac{m_j}{2\pi k T_j}\right)^{\frac{3}{2}} \exp\left(-\frac{m_jv^2}{2kT_j}\right)
\end{equation}
where $n_{0,j}$ is the reference density (in quasi-neutral state) of the species $j$. Generally, the distribution function is unknown and assumption for MW VDF is usually made. In case of the presence of an electric field, the VDF can be modified to account for the conservation of energy
\begin{equation}
    \frac{1}{2}mv^2 + q \phi = \mathrm{const.}
\end{equation}
that is the exchange between kinetic energy of a particle and energy of an electric field. The MW VDF~\eqref{eq:22_MW_VDF} can be then rewritten as~\cite{bittencourt}
\begin{equation}\label{eq:22_MW_VDF_phi}
    f_j^{\mathrm{MW}}(v) = n_{0,j} \left( \frac{m}{2\pi k T_0}\right)^{\frac{3}{2}} \exp\left(-\frac{\frac{1}{2}m_jv^2 + q_j\phi}{kT_0}\right)
\end{equation}
Such modification is commonly used for the description of a plasma sheath~\cite{Alvarez-Laguna_2020}, a region with strong electric fields. The added potential field further modifies all macroscopic quantites that can be obtained from the VDF. For instance, now the random flux, compared to~\eqref{eq:21_flux}, is calculated as
\begin{equation}\label{eq:21_flux_calc}
\begin{gathered}
    J^{\mathrm{MW}}_j = 4\pi \int^{\infty}_{0} f_j^{\mathrm{MW}}(v) v^2 \, \dd v = \\ 
    \frac{1}{4} n_{0,j} \sqrt{\frac{8 k T_j}{\pi m_j}} \exp\left(-\frac{q_j \phi}{k T_j} \right)
\end{gathered}
\end{equation}
where $n_j(\phi) = n_{0,j} \exp\left(-\frac{q_j \phi}{k T_j} \right)$ is the Boltzmann relation. Suppose that for electrons we have $q_j = -e$, which then results in a decrease in density in the vicinity of negatively charged bodies. Thus when electrons initiate to charge a solid body, the flux of electrons should decrease in time due to the decreasing potential on the solid body until the flux of positive charges balance each other.

\subsubsection{The Kappa distribution function}

\begin{figure}[ht]
    \centering
    \includegraphics[width=1.\columnwidth,trim={0.4cm 0cm 0.4cm 0},clip]{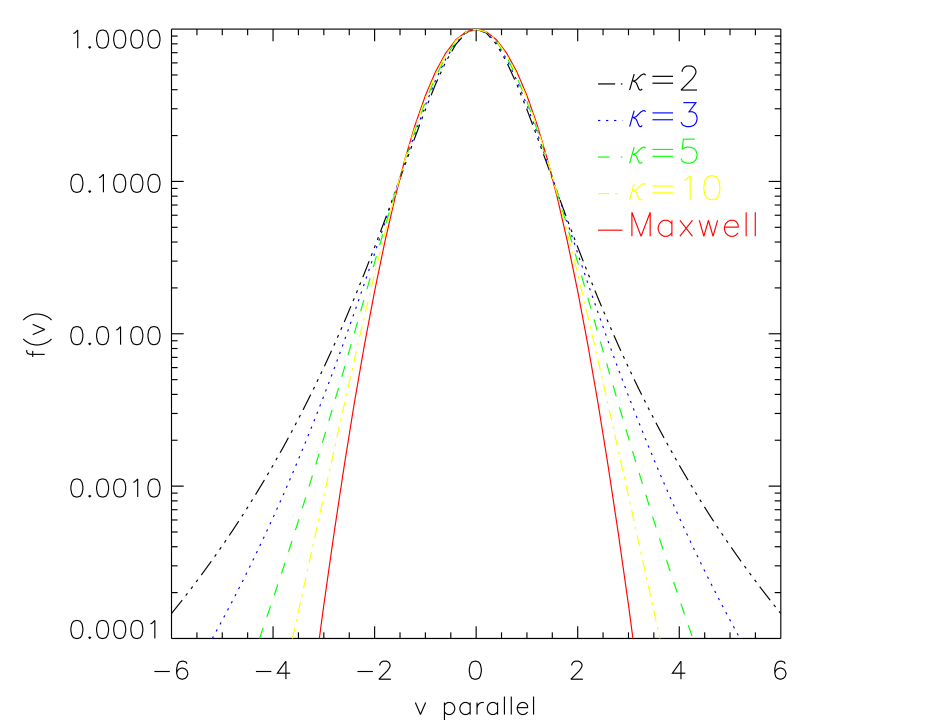}
    \caption{Comparison between the MW VDF and the Kappa VDF in high energy tail. Taken from~\cite{Pierrard}}
    \label{fig:22_kappa_distr}
\end{figure}

The assumption for MW VDF for plasma particles is usually valid in laboratory plasmas due to a sufficient amount of collisions to reach a thermal equilibrium~\cite{piel,bittencourt}. The scenario is a bit different for space plasmas where collisions are mostly absent, having enourmous mean free path at high altitudes in Earth's outer atmosphere~\cite{Nicolaou_2018,Livadiotis_2010}. Due to this fact the distribution of particles tend to deviate from the thermal equilibrium. Such deviation can be described by the Kappa law distribution function that reads~\cite{Guo_2021}
\begin{equation}
\begin{gathered}\label{eq:22_kappa_distribution}
        f_j^{\kappa}(v) = n_{0,j} \left( \frac{m_j}{2\pi k T_j (\kappa-\frac{3}{2})}\right)^{\frac{3}{2}} \frac{\Gamma(\kappa+1)}{\Gamma(\kappa-\frac{1}{2})} \\ 
        \hfill \left(1+\frac{\frac{1}{2}m_jv^2 + q_j\phi}{kT_j(\kappa-\frac{3}{2})}\right)^{-\kappa-1}
\end{gathered}
\end{equation}
where $\kappa \geq 2$ is a constant that characterizes the order of deviation. For $\kappa \to \infty$ the VDF~\eqref{eq:22_kappa_distribution} collapses into MW VDF with electric field~\eqref{eq:22_MW_VDF_phi}. Kappa distributions have been studied mostly in plasmas in the magnetosphere and solar winds, consisted of MW core and Kappa halo, although some deviations have also been reported in the ionosphere, especially in auroras for electrons. The main difference between the MW and Kappa VDFs is the suprathermal tail, a high energy tail that is more populated for Kappa VDF. The comparison can be seen in~\autoref{fig:22_kappa_distr}.

\begin{figure}[ht]
    \centering
    \includegraphics[width=0.8\columnwidth,trim={0.4cm 0.3cm 0.4cm 0},clip]{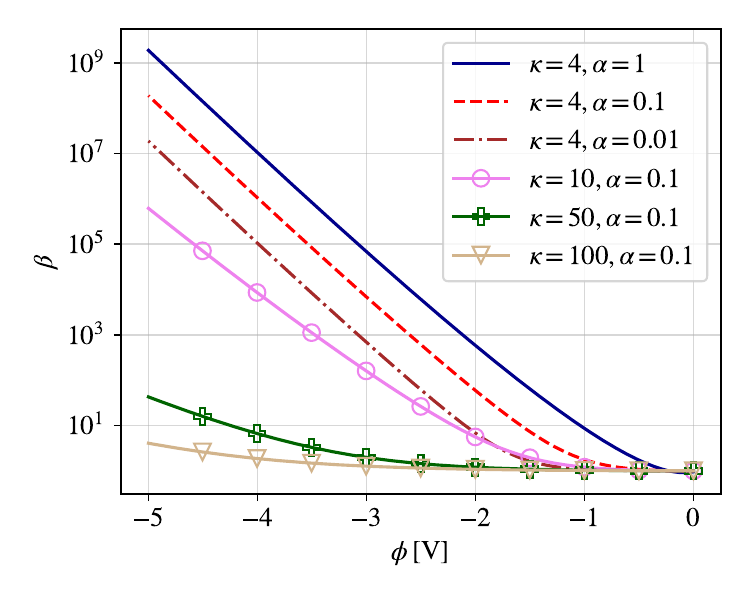}
    \vspace{-4mm}
    \caption{Ratio between the fluxes from Kappa distribution and MW distribution. Lines with markers differ in the $\kappa$ constant, while lines without markers differ in the VDF ratio $\alpha$.}
    \label{fig:22_beta_ratio}
\end{figure}

 In terms of space platform charging, there will be more electrons with sufficient energy to penetrate the potential barrier in the vicinity of solid body's surface. This can be shown by evaluating the random flux from the Kappa VDF as in~\eqref{eq:21_flux_calc}, by replacing the VDF with $f_j^\kappa$. This results in
\begin{equation}
\begin{gathered}\label{eq:22_kappa_flux}
        J_j^\kappa(\phi) = \frac{1}{4}n_{0,j} \sqrt{\frac{8 k T_j}{\pi m_j}} \frac{\Gamma(\kappa+1)}{\Gamma(\kappa-\frac{1}{2})} \frac{\sqrt{\kappa-\frac{3}{2}}}{\kappa(\kappa-1)} \\
        \hfill \left(1 + \frac{q_j\phi}{kT_j(\kappa-\frac{3}{2})}\right)^{-\kappa+1}
\end{gathered}
\end{equation}
To show the differences between the random fluxes~\eqref{eq:21_flux_calc} and~\eqref{eq:22_kappa_flux}, we employ a flux ratio $\beta$ as follows
\begin{equation}\label{eq:22_flux_ratio}
    \beta = \frac{\alpha J^\kappa+(\alpha-1)J^{\mathrm{MW}}}{J^{\mathrm{MW}}}
\end{equation}
where $\alpha$ is the portion of particles with Kappa distribution in the plasma. It can be seen as a mixture of particles having MW distribution and particles with Kappa distribution. In~\autoref{fig:22_beta_ratio}, we show the flux ratio for electron particles, varying both $\kappa$ and $\alpha$. It is interesting to see that the flux of MW electrons becomes negligible for few volts of negative potential. This means that when a surface is being charged, high energy electrons with Kappa distribution still have a sufficient energy to charge the surfaces to greater negative potential, which can lead to a different nature of space platform charging even for a small fraction of Kappa particles in a plasma. 

\subsubsection{The validity of the MW distribution in space plasma}
Any state of a collisional system of particles eventually end up in a state of thermodynamic equilibrium that is described by the MW VDF. In this part, we review a simple analysis to determine if electrons can be described by MW VDF in the ionosphere. 

Electrons can collide with all particles in the ionosphere during which a momentum between particles is exchanged. The question is how often and how important are those in order to reach the thermal equilibrium. In literature, it has been shown that the collision that plays a role in ionosphere is the Coloumb interaction with other electrons, thus electron-electron collisions~\cite{ASHIHARA19741201}. We seek to estimate at which distances can electrons thermalize with respect to the typical lengths of solid bodies $\sim 1\: \mathrm{m}$ in the ionosphere. The mean free path can be calculated as
\begin{equation}
    l = \frac{1}{n_{\rm e} \sigma_{\mathrm{m}}}
\end{equation}
where $n_{\rm e}$ is the density of colliding electrons and $\sigma_{\mathrm{m}}$ is the cross section for momentum exchange during Coloumb-like collisions of electrons. It has been shown that for Coloumb collision with Debye shielding, the momentum exchange cross section reads~\cite{bittencourt}
\begin{equation}
    \sigma_{\mathrm{m}} = 4 \pi b_0^2 \ln\left( \frac{\lambda_{\mathrm{D}}}{b_0}\right)
\end{equation}
where $b_0 = \frac{e^2}{12\pi\varepsilon_0 k T}$ is the impact parameter for high angle deflection (which assumed MW distribution). By putting in the ionosphere plasma parameters from~\autoref{tab:41_plasma_conditions}, one finds out that the mean free path for electron-electron collision is approximately $l = 1$ km. Which is much larger than the typical length scale of satellites. Other work reports even 10 km using computational tools~\cite{tsoumpariotis}. Thus this can conclude that if there is a population of electrons with Kappa distribution (or genreally non-equilibrium distribution), than it is possible that the MW VDF might not be completely precise for the description of ionosphere electrons.
\subsection{Ionospheric aerodynamics of satellites}\label{subsec:23}
The drag in orbits comes mainly from the neutral particles composing the residual atmosphere and can be evaluated as~\cite{lafleur}
\begin{equation}
    \vb{F}_{\mathrm{n}} = -\frac{1}{2} \rho_{\mathrm{n}} u \vb{u} A C_{\mathrm{D,n}}
\end{equation}
where $\rho_{\mathrm{n}}$ is the mass density of a gas, $\vb{u}$ is the velocity of a solid body in orbit relative to the flow of atmosphere gas, $A$ is the cross section of the body exposed to the flow and $C_{\mathrm{D,n}}$ is the drag coefficient of the body. When a solid body flies through the ionosphere, an additional drag on the body is introduced due to the presence of the plasma. This contribution is mostly due to the heavy ions since electrons have small mass. The drag of ions as charged particles is then~\cite{lafleur}
\begin{equation}
    \vb{F}_{\mathrm{c}} = -\frac{1}{2} \rho_{\mathrm{i}} u \vb{u} A C_{\mathrm{D,c}}
\end{equation}
where $\rho_{\mathrm{i}}$ is the ion mass density, $C_{\mathrm{D,c}}$ is the drag coefficient of the body due to the charged particles. The coefficient $C_{\mathrm{D,c}}$ differs from $C_{\mathrm{D,n}}$ due to the effect of plasma--body interaction and platform charging. 

Spacecraft charging leads to a plasma phenomenon known as the plasma sheath (sometimes called as Debye sheath) formation~\cite{piel,bittencourt,Alvarez-Laguna_2020}. This sheath is characterized as a charged region of a disturbed plasma that naturally arises around bodies immersed in a plasma. If then charged particles fly in the vicinity of the charged body, they will interact with the plasma sheath. This mechanism is interesting to study in terms of the drag force, since there will be an indirect momentum exchange between the particle and the solid body through the electrostatic forces. It can be regarded as if the charged particle interacts with the electric potential field of the solid body that can be seen as a big charged particle. An illustration of such mechanism is presented in~\autoref{fig:23_flyby_sheath}. The width of the sheath is characterized by the Debye length defined in~\eqref{eq:21_debye_length}.

\begin{figure}[ht]
    \centering
    \includegraphics[width=1.\columnwidth,trim={0.4cm 0.3cm 0.4cm 0},clip]{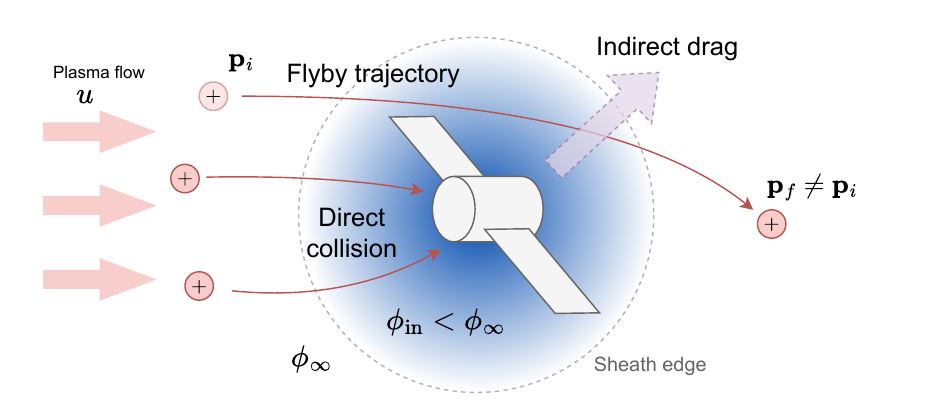}
    \vspace{-4mm}
    \caption{Illustration of indirect momentum exchange mechanism in ionosphere flows. The potential $\phi_{\infty}$ is the reference potential of undisturbed plasma, $\phi_{\mathrm{in}}$ is the potential field distribution inside the sheath for negatively charged body. Ion in flyby trajectory exchanges momentum with the solid body.}
    \label{fig:23_flyby_sheath}
\end{figure}

At this point we have two mechanisms that contribute to the drag force of charged particles, the direct collision and indirect collison (flyby trajectories). The drag coefficient of charged particles $C_{\mathrm{D},c}$ essentially accounts for both of these effects, although they could be divided into two components
\begin{equation}\label{eq:23_drag_coeff_sep}
    \vb{F}_{\mathrm{c}} = \vb{F}_{\mathrm{c}}^{\mathrm{d}} + \vb{F}_{\mathrm{c}}^{\mathrm{i}} = -\frac{1}{2} \rho_{\mathrm{i}} u \vb{u} A (C_{\mathrm{D,c}}^{\mathrm{d}} + C_{\mathrm{D,c}}^{\mathrm{i}})
\end{equation}
where the exponent indeces $\mathrm{d}$ and $\mathrm{i}$ are the direct and indirect cases respectively. This notation can be used for the calculation of individual drag coefficients. In terms of the indirect force, one could deduce that the width of the plasma sheath plays a vital role, thus depending on the state of the plasma according to~\eqref{eq:21_debye_length}.

The indirect drag force can be estimated from the analysis of momentum equations for plasma particles. The work~\cite{capon2} showed that for plasma interacting with a body, the resulting force is evaluated as
\begin{equation}
    \vb{F}_{\mathrm{c}} = \underbrace{-\int_{S} \rho_{\mathrm{i}} \vb{u} (\vb{u} \cdot \hat{\vb{n}}) \, \dd S}_{\vb{F}_{\mathrm{c}}^{\mathrm{d}}} + \underbrace{\int_{S} \overline{\vb{T}} \cdot \hat{\vb{n}} \dd S}_{\vb{F}_{\mathrm{c}}^{\mathrm{i}}}
\end{equation}
where $S$ is the surface of the solid body, $\hat{\vb{n}}$ is an unit vector normal to the surface in outward direction and $\overline{\vb{T}}$ is known as the Maxwell stress tensor 
\begin{equation}\label{eq:23_maxwell_stress}
    \overline{\vb{T}}_{ij} = \varepsilon_0 \left(E_i E_j - \frac{1}{2} \delta_{ij} E^2\right) + \frac{1}{\mu_0} \left(B_i B_j - \frac{1}{2} \delta_{ij} B^2\right)
\end{equation}
where $E_i$ is the $i$-th component of electric intensity, $B_i$ is the component of the magnetic field and $\mu_0$ is the vacuum permeability. In this work, we neglect any induced magnetic field due to the currents and we neglect the contribution of Earth's magnetic field, thus $\vb{B} \approx \vb{0}$. The Maxwell stress tensor describes the applied pressure of electromagnetic field on a surface and thus the net exchange momentum due to the deformation of the plasma sheath. In case of stationary plasma, the plasma sheath remains symmetrical and, as a result, the indirect forces will cancel out. The illustration of the indirect force is shown in~\autoref{fig:23_flyby_sheath}. In the case of satellite flights, a deformation of the plasma sheath is observed.

\begin{figure}[ht]
    \centering
    \includegraphics[width=1.\columnwidth,trim={0.4cm 0.3cm 0.4cm 0},clip]{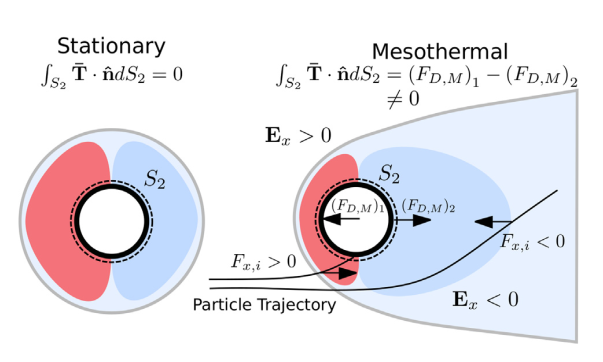}
    \vspace{-4mm}
    \caption{Illustration of plasma sheath deformation linked with indirect force. Picture taken from~\cite{capon2}.}
    \label{fig:23_flyby_sheath}
\end{figure}

\section{Particle-In-Cell Simulations}\label{sec:3}
\subsection{Simulation of plasma dynamics}\label{subsec:31}
To describe the dynamics of plasma with solid body interactons we utilized one of the commonly used particle methods known as Particle-In-Cell (PIC) method. This method introduces the concept of a superparticle; a numerical particle that consists of many particles of the same kind. This reduces the number of real particles for given density at the cost of increased statistical noise~\cite{lapenta,hockney2021computer}. In this work, the ratio between real particles and superparticles is denoted as $F_{\mathrm{num}}$. In this case, the mass and charge of a superparticle for a given $F_{\mathrm{num}}$ is
\begin{equation*}
    M_j = F_{\mathrm{num}} \cdot m_j, \quad Q_j = F_{\mathrm{num}} \cdot q_j
\end{equation*}
The evolution of superparticle's (from now on we continue with particle) trajectory is given by the force law in an electromagnetic field
\begin{equation}\label{eq:31_force_law}
    \vb{F} = M \dv{\vb{v}}{t} = Q (\vb{E} + \vb{v} \times \vb{B} )
\end{equation}
where $\vb{E}$ is the electric intensity and $\vb{B}$ is the magnetic field. Relativistic effects are neglected in this work. For an electrostatic case, the electric intensity is connected to the electric potential field as
\begin{equation}
    \vb{E} = - \grad{\phi}
\end{equation}
where $\phi$ can be found from~\eqref{eq:21_poisson_equation}. For this purpose, the PIC method introduces a discretized mesh of a simulation domain using cells. The Poisson equation is then solved at each node of the mesh which leads to a Finite Element Method (FEM)\cite{may}. The equation~\eqref{eq:21_poisson_equation} is then solved for each cell in the mesh as follows
\begin{equation}\label{eq:31_fem_poisson}
   \varepsilon \phi_k \int_C \grad v_k \grad v_l \dd V = \sum_{p\in C}  \int_C \delta(\vb{r}-\vb{r}_p) v_l \, \dd V
\end{equation}
where the sum is over all particles $p$ (of any species) that belong to the cell $C$. Here $v_k$ is the test function and $k,l$ are two of the nodes of cell $C$ (e.g. for triangular mesh we have $k=1,2,3$). Further $\delta(\vb{r}-\vb{r}_p)$ is a delta function that represents a point particle. The integral of a delta function multiplited by the test function represents a weighing assignment to the nodes. The test functions are defined in a way that
\begin{equation}
    v_k(\vb{r_l}) = 
    \begin{cases}
        1, &\qquad \textrm{if} \quad k=l \\
        0, &\qquad \textrm{if} \quad k\neq l
    \end{cases}
\end{equation}
When accounted for all cells in the mesh, we end up with a linear system with size given by the number of nodes. Here $\int_C \grad v_k \grad v_l \, \dd V$, for given $k,l$ over all cells, is a matrix,  that needs to be inverted to obtain the solution of the potential field $\phi_k$ in all nodes. The order of the numerical solution is given by the order of the test function (represented by e.g. linear functons, polynomials, etc.)

When the solution is obtained, the potential field is interpolated back to the particles which can be the used for the force law~\eqref{eq:31_force_law} while simulatenously solving following equation
\begin{equation}\label{eq:31_deriv_position}
    \dv{\vb{r}}{t} = \vb{v}
\end{equation}
where $\vb{r}$ is the position of the particle.
To obtain the new velocity and position, the equations are discretized in time. Some of the typical \textit{push} schemes for the solution can be found in~\cite{lapenta,zenitani}.

The solutions to equations~\eqref{eq:21_poisson_equation}, \eqref{eq:31_force_law}, \eqref{eq:31_deriv_position}  in discretized environment need to satisfy specific stability conditions depending on the nature of the particles. For the time discretization, the length of a time step, denoted as $\Delta t$, needs to be chosen in a way that the particle does not travel more than one cell width
\begin{equation}\label{eq:21_time_restriction}
    \Delta t < \frac{\Delta C}{\vb{v}}
\end{equation}
where $\Delta C$ is the average cell size. For plasma particles, the speed is usually taken as the mean thermal speed $\overline{u}$. Furthermore, in all simulations, the Debye sheath must be resolved for accurate results. The cell size is limited accordingly
\begin{equation}\label{eq:21_debye_restriction}
    \Delta C < 0.5 \lambda_{\mathrm{D}}
\end{equation}
where the constant 0.5 is usually chosen~\cite{capon3}. Lastly plasma oscillations at the electron plasma frequency $\omega_{\mathrm{p}}$ need to be accounted for
\begin{equation}
    \Delta t < \omega_{\mathrm{p}} = \sqrt{\frac{n e^2 }{m \varepsilon_0}}
\end{equation}

\subsection{Development of the Boltzmann electron fluid model}\label{subsec:32}
In most situations, electrons introduce heavy constraints to the time stepping due to their low mass (electrons move much faster than heavy ions) according to the mean thermal speed
\begin{equation}
    \overline{u}_{\mathrm{e}} = \sqrt{\frac{8 k T_{\mathrm{e}}}{\pi m_{\mathrm{e}}}}
\end{equation}
For that reason, it is a common practice to replace electron particles with a fluid description. This is achieved by assuming Maxwell-Boltzmann VDF~\eqref{eq:22_MW_VDF_phi} for electrons and by integrating over velocity space to obtain
\begin{equation}\label{eq:32_boltz_relation}
    n_{\mathrm{e}} (\phi) = n_{0,\mathrm{e}} \exp\left( \frac{e\phi}{k T_{\mathrm{e}}}\right)
\end{equation}
Such relation assumes that electrons have MW distribution in electrostatic field~\cite{chen1984introduction}. The discretized Poisson equation~\eqref{eq:31_fem_poisson} is then rewritten after substituing for the electron density~\eqref{eq:32_boltz_relation} on the right hand side
\begin{equation}\label{eq:32_boltz_poisson}
\begin{gathered}
       \varepsilon \int_C \phi_k \grad v_k \grad v_l \dd V + e n_{0,\mathrm{e}} \int_C  \exp\left( \frac{e \phi_k}{k T_{\mathrm{e}}}\right) v_k v_l \, \dd V \\ = \sum_{p\in C}  \int_C \delta(\vb{r}-\vb{r}_p) v_l \, \dd V
\end{gathered}
\end{equation}
where we expanded the exponential in terms of the test functions. The equations~\eqref{eq:32_boltz_relation}, \eqref{eq:32_boltz_poisson} eventually represent the Boltzmann electron fluid model, and in general picture we end up with Hybrid PIC method -- coupling particle and fluid methods, since the sum on the right hand side of~\eqref{eq:32_boltz_poisson} still accounts for simulated particles, while leaving out the electrons. Such result unfortunately becomes non-linear in terms of $\phi_k$ but with the gread advantage of removing all the electron time scales that bear a heavy constraint. The solution to equation~\eqref{eq:32_boltz_poisson} requires iterative numerical solvers for non-linear equations. 
\subsection{Charging of surfaces}
When particle arrives to the surface of a solid body, either it gets reflected, attaches or reacts on the surface. In case of attached charged particles, we accumulate net charge on the surface of a solid body. This effect is accounted for by the $\delta$ function on the right hand side of~\eqref{eq:31_fem_poisson}. When we integrate the equation in a cell that is in touch with a surface of the solid body, the attached particles appear in the sum that has been collected during a previous time step when the equation is solved. Depending on the material, the attached particles can be either conducted along the surface to reach a uniform net charge over the solid surface, or it can be isolated in a place where collision occured. In that case the body is then non-conductive. The latter is assumed in this work.

In case of Hybrid PIC with Boltzmann electrons, the implementation is a bit different. Instead of discrete accomodation of single charges on a surface, we are required to use a macroscopic quantity obtained from distribution functions. The accumulated charge $Q^{\rm acc}$ due to the Boltzmann electrons on arbitrary surface in some time $t$ is then evaluated using the random flux as follows
\begin{equation}
    Q^{\rm acc} = - e \int_{t_0}^t \int_{S}  J^{\mathrm{MW}}_{\mathrm{e}} \, \dd S \dd t
\end{equation}
where $S$ is the surface being charged, $t_0$ is initial time and $J^{\mathrm{MW}}_{\mathrm{e}}$ is the Maxwellian random flux defined in~\eqref{eq:21_flux_calc}. One can notice that the accumulated charge can change if we used the random flux $J^{\kappa}$~\eqref{eq:22_kappa_flux} from the Kappa distribution. Therefore, for Hybrid PIC, we obtain final form of the Poisson equation
\begin{equation}\label{eq:33_charge_poisson}
\begin{gathered}
       \varepsilon \int_C \phi_k \grad v_k \grad v_l \dd V + e n_{0,j} \int_C  \exp\left( \frac{e \phi_k}{k T_{\mathrm{e}}}\right) v_k v_l \, \dd V  = \\ \sum_{p\in C}  \int_C \delta(\vb{r}-\vb{r}_p) v_l \, \dd V + Q^{\rm acc}_l
\end{gathered}
\end{equation}
where we integrate through surface of a solid body in $Q^{\rm acc}$ that belongs to the cell $C$.

\subsection{Drag force estimation on solid body}\label{subsec:34}
For the numerical estimation of drag force, we need to evaluate both direct forces from particle collisions with solid body and the indirect force which is described by the Maxwell stress tensor~\eqref{eq:23_maxwell_stress}. For the direct force, we directly calculate the force as the change of momentum of a particle (works for both neutral and charged particles) after it hits a surface during a time step $\Delta t$. This is evaluated as
\begin{equation}
    \vb{F}_{\mathrm{b}}^{\mathrm{d}} = \frac{1}{\Delta t} \left(\vb{p}_i - \vb{p}_f\right)
\end{equation}
where $\vb{F}_{\mathrm{b}}$ is the direct force acting on a body, $\vb{p}_i = M \vb{v}_i$ is the momentum before collision and $\vb{p}_f = M \vb{v}_f$ is the momentum after collision. For attachment it is clear that $\vb{v}_f = \vb{0}$.

For the indirect force, we simply integrate the Maxwell stress tensor over the whole surface of a solid body 
\begin{equation}
    \vb{F}_{\mathrm{b}}^{\mathrm{i}} = \sum_{C} \int_{S \in \partial C} \overline{\vb{T}} \cdot \hat{\vb{n}} \dd S
\end{equation}
where $\vb{F}_{\mathrm{b}}^{\mathrm{i}}$ is the indirect force acting on a body and the sum goes through all cells in contact with a solid body.
\subsection{PANTERA software}\label{subsec:34}
\begin{figure}[ht]
    \centering
    \includegraphics[width=1.\columnwidth,trim={0.4cm 1.0cm 0.4cm 1cm},clip]{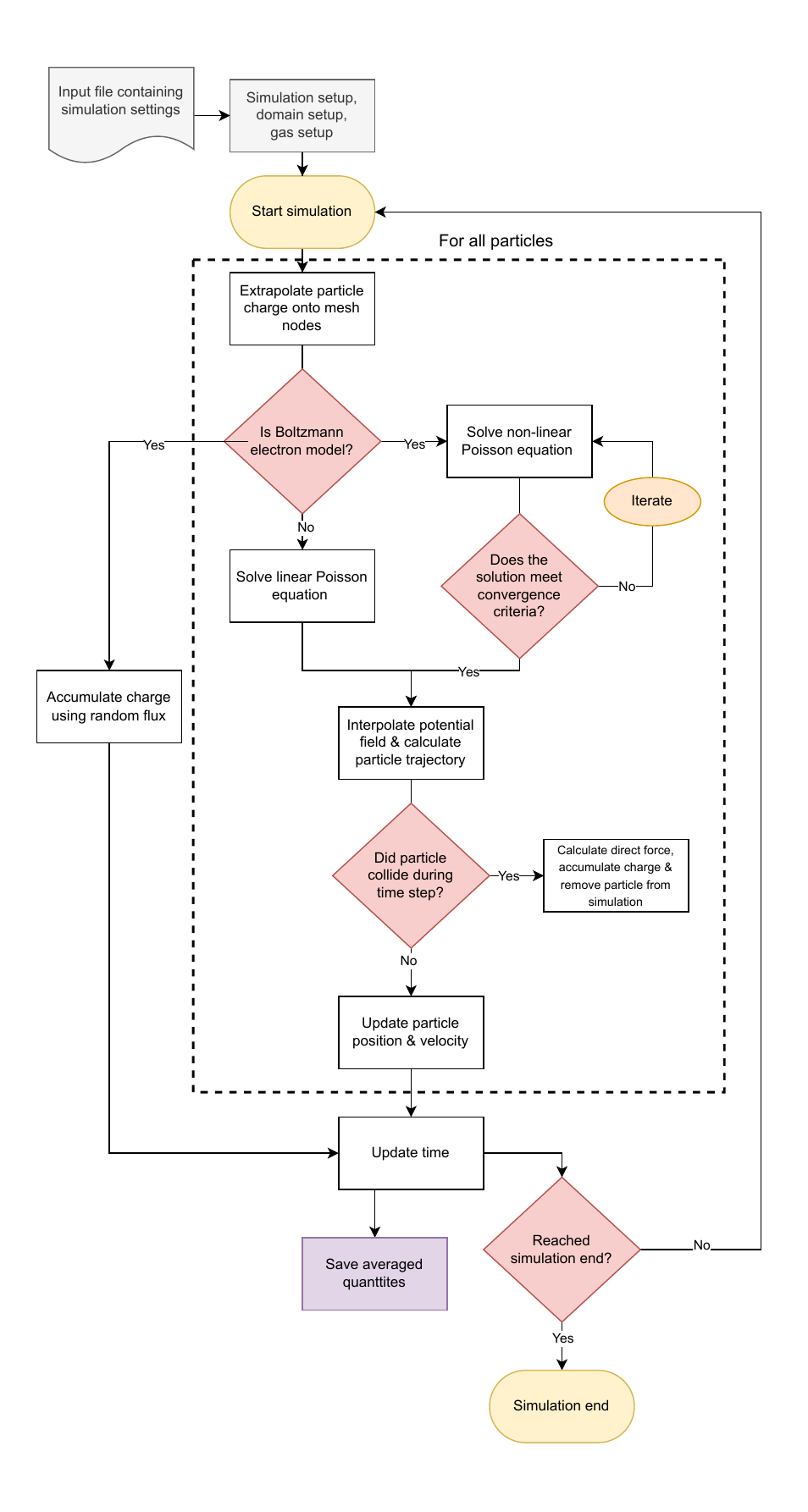}
    \vspace{-4mm}
    \caption{Illustration of algorithm in PANTERA for space platform charging.}
    \label{fig:35_pig_algorithm}
\end{figure}

In this work we utilize the PArticle Numerical Tool for non-Equilibrium Reacting Aerodynamics (PANTERA) code developed in-house at the von Karman Institute for the simulation of plasma dynamics~\cite{pantera}. It implements the PIC method that uses the Boris scheme for the trajectory update of particles. The field equations are solved using the FEM method~\eqref{eq:31_fem_poisson}. In this work, we have implemented the Boltzmann fluid model, the Kappa distribution and the drag force estimation in PANTERA and tested on verification cases that are shown in~\autoref{sec:4}. The illustration of a simulation algorithm used in this work can be seen in~\autoref{fig:35_pig_algorithm}. PANTERA itself does not have mesh generator, an SU2 file of the mesh needs to be passed to the input to successfully generate the computational domain. Lastly, the PANTERA software utilizes the PETSc library, a mathematical tool set for the solution of linear and non-linear systems of equations~\cite{petsc}. In this work, a non-linear solver from PETSc is used for the solution of the non-linear Poisson equation.

\section{Verification of implemented models}\label{sec:4}
This section is devoted to thorough verification of the implementations. In the following section~\ref{subsec:41} we employ a 1D test case, which consists of a formation of plasma sheath on a wall, that will allow us to test the implemented non-linear solver for Poisson equation and the indirect drag force computation on a solid body. Section~\ref{subsec:42} is focused on a verification of direct drag force calculation using a simple sphere that interacts with flowing neutral particles.

\subsection{1D Plasma Sheath}\label{subsec:41}
\begin{figure}[ht]
    \centering
    \includegraphics[width=0.8\columnwidth,trim={0.4cm 0.3cm 0.4cm 0},clip]{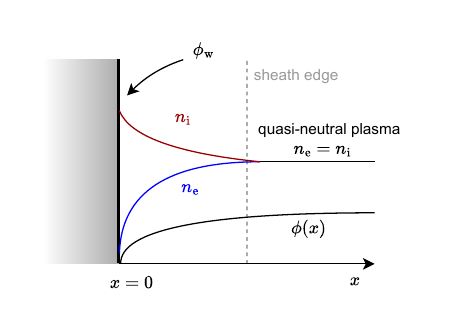}
    \vspace{-4mm}
    \caption{Computational domain for the simulation of 1D plasma sheath.}
    \label{fig:41_wall_sheath}
\end{figure}

Suppose a 1D case on $x$ axis that is bounded by wall from one direction as shown in~\autoref{fig:41_wall_sheath}. The Poisson equation can be then written as
\begin{equation}\label{eq:41_poisson}
    \varepsilon \dv[2]{\phi(x)}{x} = e \left(n_{\mathrm{e}}(x) - n_{\mathrm{i}}(x)\right)
\end{equation}
where $n_e(x)$ and $n_{\mathrm{i}}(x)$ is the electron and ion density respectively. A uniform permitivity is assumed. The density functions are substituted with Boltzmann relations~\cite{bittencourt}
\begin{equation}\label{eq:41_boltzmann_relations}
    n_{\mathrm{e,i}} = n_0 \exp \left[ \frac{\pm e \phi(x)}{k T_0}\right]
\end{equation}
where $n_0$ is the reference density of quasi-neutral plasma, $T_0$ is the reference temperature. The plus sign is accounted for electrons and minus sign for ions. By substituing the density functions back to the Poisson equation~\eqref{eq:41_poisson} we arrive to second order differential equation with following boundary conditions~\cite{bittencourt}
\begin{align}
    \epsilon \dv[2]{\phi}{x} &= 2 e n_0 \sinh\left(\frac{e\phi}{kT_0}\right)\label{eq:41_differential_equation} \\
    \phi(0) &= \phi_{\mathrm{w}}, \quad \phi(\infty) = 0 
\end{align}
where $\phi_{\mathrm{w}}$ is the wall potential. The solution to~\eqref{eq:41_differential_equation} is 
\begin{align}
    \phi(x) &= -4 \frac{k T_0}{e} \arctanh\left(C e^{-\frac{\sqrt{2}}{\lambda_{\mathrm{D}}}x}\right) \label{eq:41_poisson_solution} \\
    C &= \tanh\left(- \frac{e \phi_{\mathrm{w}}}{4k T_0}\right) \label{eq:41_constant_wall}
\end{align}
where $C$ is a constant found from the boundary condition at the wall ($x=0$). For the case of platform charging verification, we account for the change of potential field in time by simply introducing time varying wall potential governed by the change of surface charge at the wall due to the thermal flux of electrons (we neglect the ion thermal flux) accordingly 
\begin{equation}
    \dv{\sigma_{\mathrm{w}}}{t} = - \frac{1}{4} e n_0 \underbrace{\sqrt{\frac{8 k T_0}{\pi m_{\mathrm{e}}}}}_{\overline{u}_{\mathrm{e}}} \exp(\frac{e\phi_{\mathrm{w}}}{kT_0})
\end{equation}
where $\overline{u}_{\mathrm{e}}$ is the mean speed obtained from Maxwell-Boltzmann velocity distribution.
From the jump condition at the boundary (with $\vb{E} = 0$ for $x<0$) we come to a cubic equation with unique solution
\begin{equation}\label{eq:41_1q}
    \phi_{\mathrm{w}}(t) = - \frac{2kT_0}{e}\ln B(t)
\end{equation}
where $B(t)$ is function evaluated as
\begin{align}
    B(t) &= \left(\frac{3}{2}D + \sqrt{\frac{9}{4}D^2 + 1}\right)^\frac{1}{3} + \left(\frac{3}{2}D - \sqrt{\frac{9}{4}D^2 + 1}\right)^\frac{1}{3} \label{eq:41_2q}\\
    D &= a_0 + \sqrt{2} \frac{\overline{u}_{\mathrm{e}}}{\lambda_{\mathrm{D}}} t \label{eq:41_3q}\\
    a_0 &= \exp(- \frac{e \phi_{\mathrm{w}}(0)}{2kT_0}) + \frac{1}{3}\exp(- \frac{3 e \phi_{\mathrm{w}}(0)}{2kT_0}) \label{eq:41_4q}
\end{align}
One can notice that the expression in~\eqref{eq:41_3q} gives us the characteristic charging time $\tau = \lambda_{\mathrm{D}}/\overline{u_{\mathrm{e}}}$ for electrons. The equations~\eqref{eq:41_poisson_solution},\eqref{eq:41_constant_wall},\eqref{eq:41_1q} give us the full analytical solution for the floating wall in 1D that can be used as verification test case. The full derivation for all steps can be found in Appendix A.

\begin{figure}[ht]
    \centering
    \includegraphics[width=0.8\columnwidth,trim={0.4cm 0.3cm 0.4cm 0},clip]{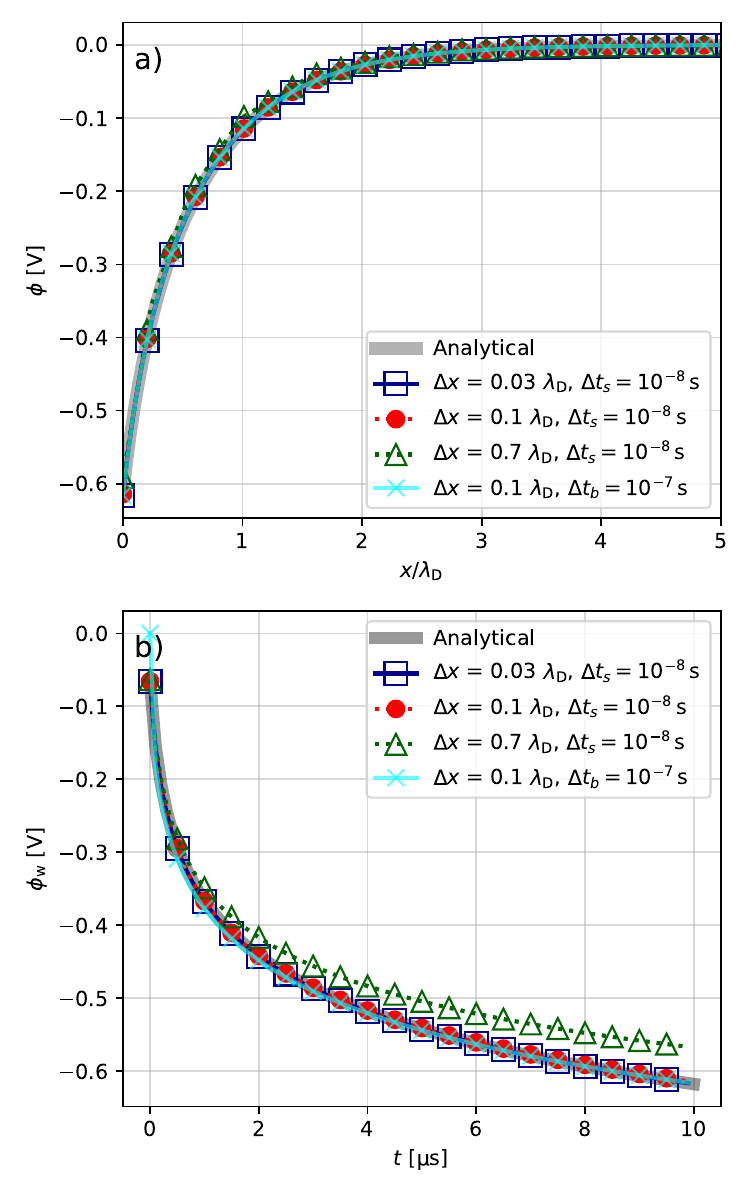}
    \vspace{-4mm}
    \caption{Solution from the simulation of 1D plasma sheath. The graph a) depicts the potential field distribution $\phi(x)$ and graph b) the wall potential $\phi_{\mathrm{w}}(t)$ in time.}
    \label{fig:41_plate_graph}
\end{figure}

We attempt to solve these equations in PANTERA, mainly the equation~\eqref{eq:41_poisson_solution} which makes the problem non-linear. Even though we do not have the exact format of a Boltzmann solver, the right hand side functions are quite similar. If we can successfully solve such equation, then we can solve an equation with Boltzmann electrons. The case domain for the potential field solution is shown in~\autoref{fig:41_wall_sheath}. The length of the domain should be at least several Debye lengths in order to resolve the potential drop region. In this case we do not simulate any particles, only the Poisson equation~\eqref{eq:41_poisson_solution} is solved at each time step to find the consistent solution for floating wall. The results from the simulation are shown in~\autoref{fig:41_plate_graph}a) for potential field $\phi(x)$ at final time step and in~\autoref{fig:41_plate_graph}b)  for the time evolution of the wall potential $\phi_{\mathrm{w}}(t)$. Both results are shown for different cell sizes and different lengths of time steps to study the accuracy of platform charging. The plasma conditions are listed in~\autoref{tab:41_plasma_conditions}. It is noticable that the solutions seem to be accurate in all cases although the solution for the wall potential with coarser mesh discretization $\Delta x = 0.7 \lambda_{\mathrm{D}}$ deviates in time.

\begin{table}[ht]
    \centering
    \caption{Plasma conditions and simulation options for 1D plasma sheath test case. Taken from~\cite{IRI} for 50°LA  50°LO, 01/06/2024 10:00.}
    \begin{tabular}{@{} >{\raggedright}p{0.5\linewidth} @{\hspace{0.5cm}} p{0.4\linewidth} @{}}
        \hline
        $n_0 \: \left[\mathrm{cm^{-3}}\right]$ & $5.62\cdot 10^{5}$ \\
        $T_0 \: \left[\mathrm{K}\right]$ & 1842 \\
        $t_{\mathrm{end}} \: \left[ \mathrm{\mu s}\right]$  & 10\\
        \hline
    \end{tabular}
    \label{tab:41_plasma_conditions}
\end{table}

\begin{figure}[ht]
    \centering
    \includegraphics[width=0.8\columnwidth,trim={0.3cm 0.3cm 0.3cm 0},clip]{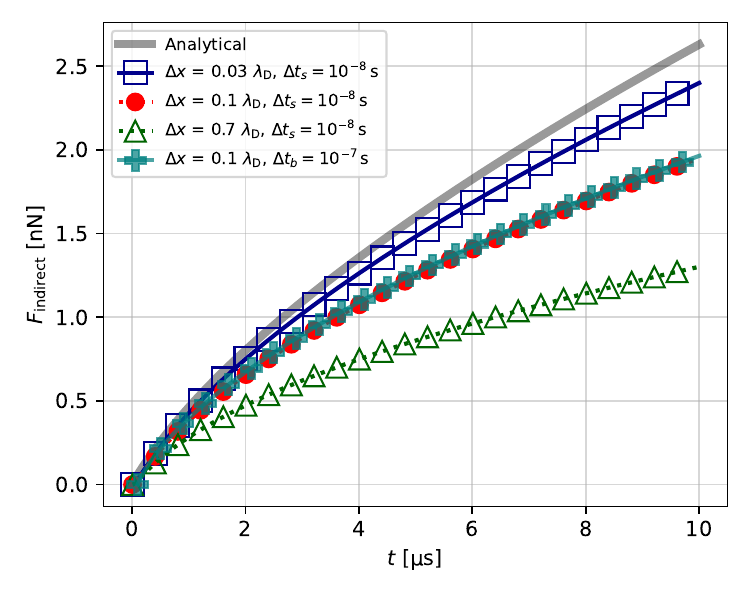}
    \vspace{-4mm}
    \caption{The calculated indirect force acting on the wall of given surface $S = \lambda_{\mathrm{D}} \times 1 \, {\mathrm{m}}$.}
    \label{fig:41_drag_force}
\end{figure}

The effects of the error can be seen in the calculation of indirect forces acting on the wall presented in~\autoref{fig:41_drag_force}. The evaluation of the force requries the gradient of the potential field to obtain the electric field $\vb{E} = - \grad{\phi}$, which introduces additional error in the discretization which is evident in the result in~\autoref{fig:41_drag_force}. Thus in order to remain accurate with respect to given error range, it requires much finer mesh. Nevertheless, it is more of importance to compare the order of the forces rather than to obtain precise estimations. To conclude the testing, we have shown that the implemented models, non-linear solver for Poisson equation and indirect drag force estimation, are successfully verified.

\subsection{Drag of neutral particles on a sphere}\label{subsec:42}
\begin{figure}[ht]
    \centering
    \includegraphics[width=0.8\columnwidth,trim={0.4cm 0.3cm 0.4cm 0},clip]{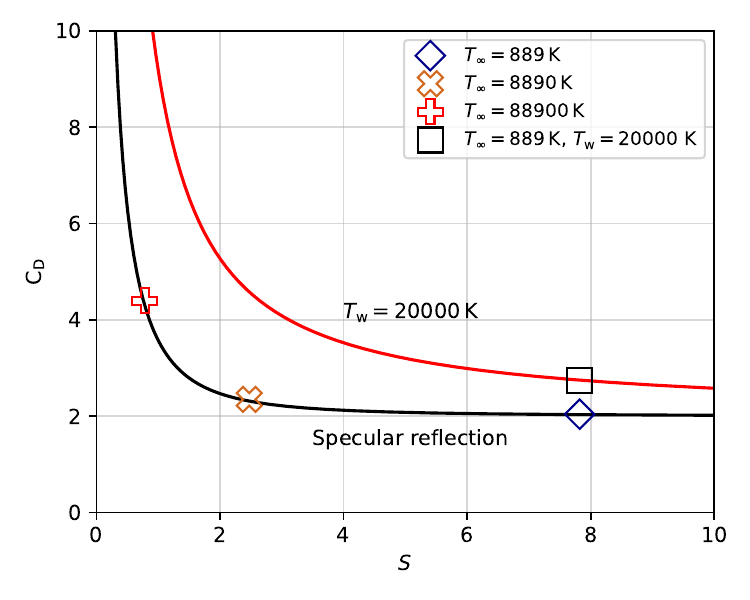}
    \vspace{-4mm}
    \caption{Comparison of obtained drag coefficients of a sphere from a simulation with the analytical solution presented in~\cite{whitfield}. The free stream velocity is kept constant for all $S = u_{\infty}/\sqrt{\frac{2 k T_{\infty}}{m}}$.}
    \label{fig:41_drag_coeff}
\end{figure}

We present additional test case to verify the implemented model for direct force calculation, that is, to compute the force due the direct impact of simulated particles on a solid body. For that purpose we compare the results to the analytical solution for the drag coefficient of the sphere interacting with flow of neutral particles given in~\cite{whitfield}(p. 25). We simulate a 2D axisymmetric domain with a sphere of diameter $d = 0.5 \: \mathrm{m}$ that is exposed to a direct flow of neutral oxygen atoms with density $n = 1.21 \cdot 10^{5} \: \mathrm{cm^{-3}}$, free stream velocity $u_{\infty} = 7.5 \: \mathrm{km \cdot s^{-1}}$ and temperature $T_{\infty}$. The domain is similar to the one shown in~\autoref{fig:43_sphere_domain} used in another test case. During the simulation, the particles hitting the surface of the solid body undergo either specular reflection or diffuse reflection (heated surface on temperature $T_{\mathrm{w}}$). The result of the simulation for both cases is shown in~\autoref{fig:41_drag_coeff}. The implemented simple model for drag force calculation is shown to be accurate for different temperature conditions.

\section{Results \& discussion}\label{sec:5}
\subsection{Ionosphere plasma--sphere interaction}\label{subsec:43}
\begin{figure}[ht]
    \centering
    \includegraphics[width=1.\columnwidth,trim={0.4cm 0.3cm 0.4cm 0},clip]{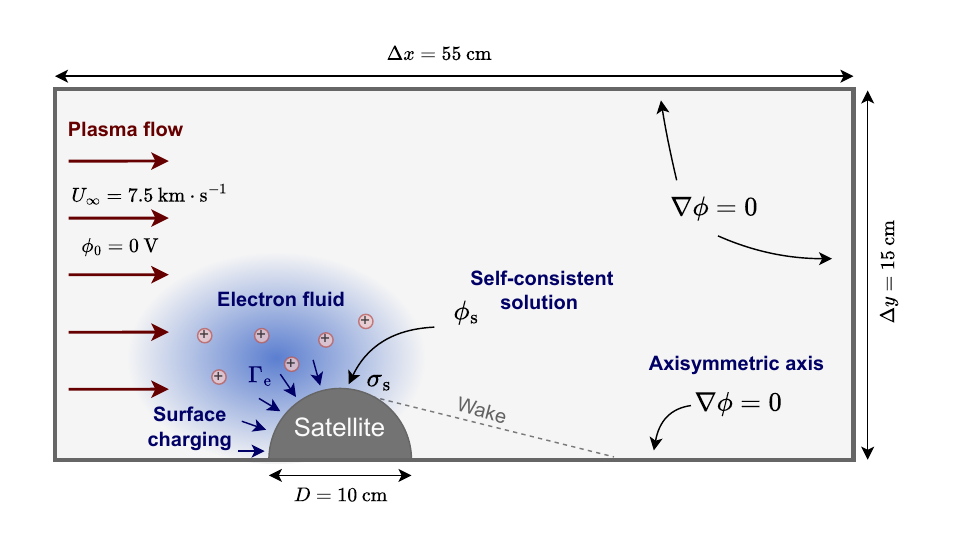}
    \vspace{-4mm}
    \caption{Illustration of a simulation domain for the ionosphere plasma--sphere interaction with Boltzmann electron fluid.}
    \label{fig:43_sphere_domain}
\end{figure}

With the Boltzmann solver implemented; coupled with direct and indirect forces calculations, we wish to test a simulation of the interaction of plasma with a sphere flying through the ionosphere. The aim here is to compare both full PIC and hybrid PIC simulation in order to test the accuracy of the Boltzmann solver for given conditions. The proposed simulation domain is depicted in~\autoref{fig:43_sphere_domain}. We inject a plasma that is composed of oxygen ions O$^+$ (i) and electrons (e). The plasma conditions are obtained from IRI 2020 for following orbit conditions $h = 250 \: \mathrm{km}$, 50° LA, 5°LO, 01/06/2024 10:00 and are shown in~\autoref{tab:43_plasma_conditions}.

\begin{table}[ht]
    \centering
    \caption{Plasma conditions for the simulation of ionosphere plasma--sphere interaction (IRI: 50°LA  50°LO, 01/06/2024 10:00).}
    \begin{tabular}{@{} >{\raggedright}p{0.5\linewidth} @{\hspace{0.5cm}} p{0.4\linewidth} @{}}
        \hline
        $n_{\mathrm{i},\mathrm{e}} \: \left[\mathrm{cm^{-3}}\right]$ & $5.62\cdot 10^{5}$ \\
        $T_{\mathrm{e}} \: \left[\mathrm{K}\right]$ & 1842 \\
        $T_{\mathrm{i}} \: \left[\mathrm{K}\right]$ & 1209 \\
        $u_{\infty} \: \left[\mathrm{km\cdot s^{-1}}\right]$ & 7.5 \\
        \hline
    \end{tabular}
    \label{tab:43_plasma_conditions}
\end{table}

\begin{figure}[ht]
    \centering
    \includegraphics[width=1.\columnwidth,trim={0.8cm 0.8cm 0.8cm 0},clip]{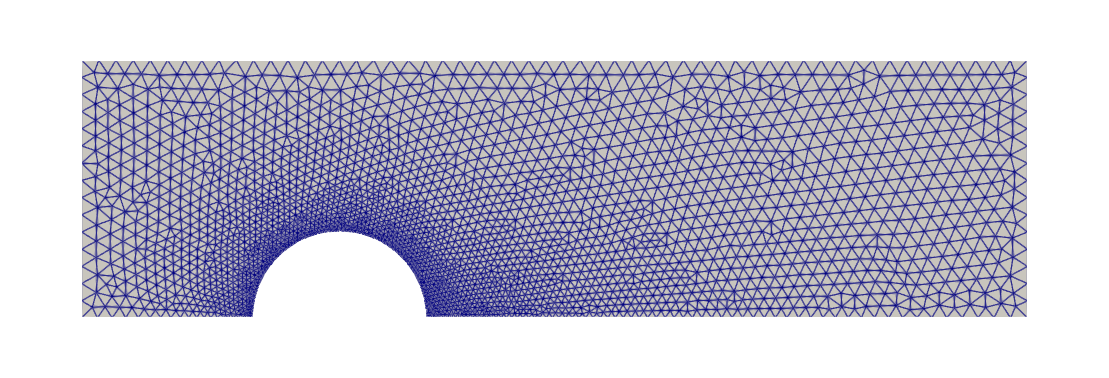}
    \vspace{-4mm}
    \caption{Meshed computational domain for plasma--sphere simulation. The mesh size converges from $\Delta C = 1 \: \mathrm{cm}$ to $\Delta C = 0.2 \: \mathrm{cm}$ in the vicinity of the sphere to resolve the Debye sheath.}
    \label{fig:43_2D_mesh}
\end{figure}

The domain consists of non-conductive sphere of diameter $D=10 \: \mathrm{cm}$, the dimensions of the box are $\Delta x = 55 \: \mathrm{cm}$ and $\Delta y = 15 \: \mathrm{cm}$. For the stability conditions~\eqref{eq:21_time_restriction},~\eqref{eq:21_debye_restriction}, for Full PIC simulation the cell size was chosen to be $\Delta C = 0.2 \: \mathrm{cm} \approx 0.5 \lambda_{\mathrm{D}}$ in the whole domain, while for Hybrid simulations, the cell size is $\Delta C = 1 \: \mathrm{cm}$ that is refined to to $\Delta C = 0.2 \: \mathrm{cm}$ in the vicinity of th solid body to resolve the potential gradients. The mesh is shown in~\autoref{fig:43_2D_mesh}. Particles of density $n_{\mathrm{i,e}} = 5.62 \cdot 10^5 \: \mathrm{cm^{-3}}$ are injected from the left boundary with velocity $u_{\infty} = 7.5 \: \mathrm{km\cdot s^{-1}}$ and at reference potential $\phi_{\infty} = 0 \: \mathrm{V}$. Neumann condition is applied to remaining boundaries. 

\subsubsection{Full vs Hybrid PIC Comparison}
\begin{figure}[ht]
    \centering
    \includegraphics[width=0.7\columnwidth,trim={0.1cm 0.4cm 0.1cm 0},clip]{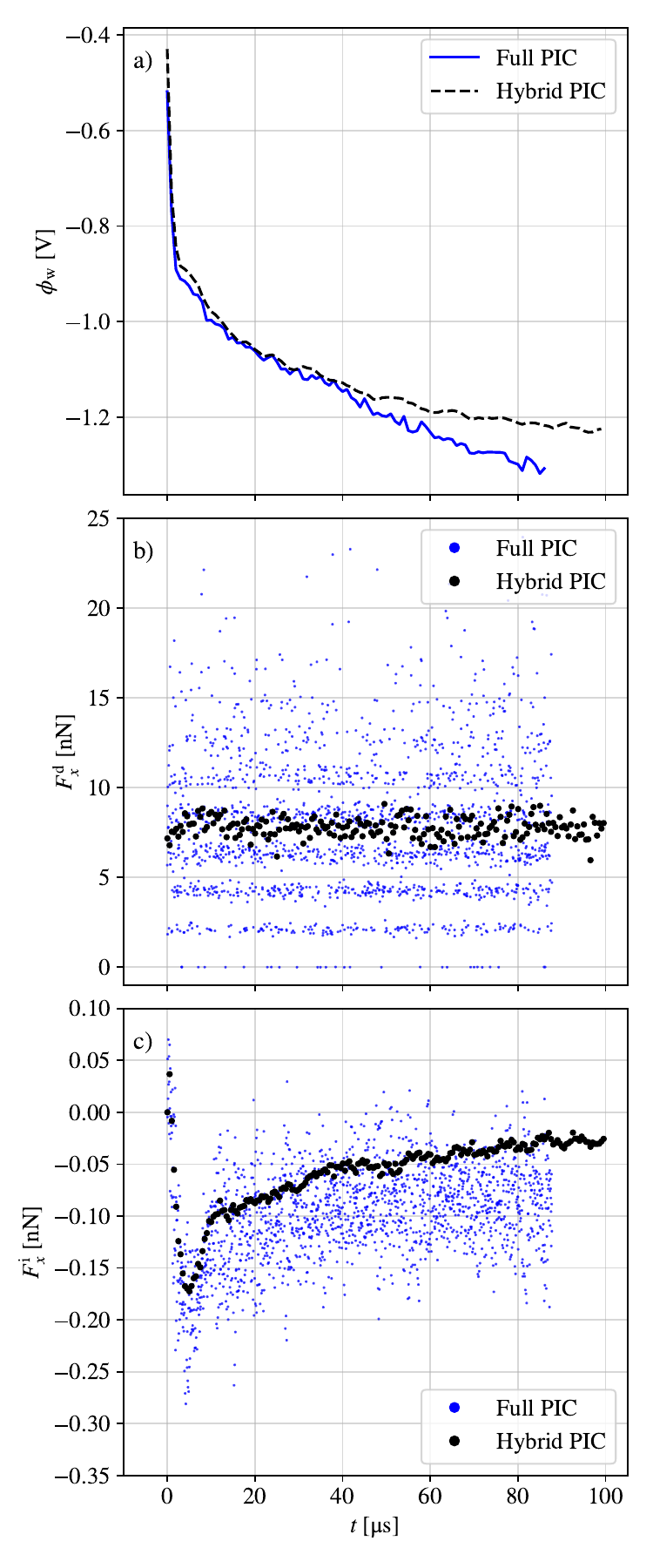}
    \vspace{-4mm}
    \caption{Results from the plasma--sphere interaction simulation for Full vs Hybrid PIC comparison. The wall potential in the shadow part of the sphere is shown a) as a function of time. In figure b) we have the direct force and in c) the indirect force.}
    \label{fig:43_short_graph}
\end{figure}

We show the first results for the comparison between fully kinetic (Full) PIC simulation where we simulate both O$^+$ and electron particles. The time step was chosen as $\Delta t = 1 \: \mathrm{ns}$ to satisfy the stability condition~\eqref{eq:21_time_restriction}, while for the Hybrid PIC simulation, the time step is set to $\Delta t = 500 \: \mathrm{ns}$. The results are shown in~\autoref{fig:43_short_graph}. The number of time steps was $10^5$ for Full PIC while for Hybrid it was $2\cdot 10^3$. In~\autoref{fig:43_short_graph}a) we have the evolution of the wall potential in shadowed region of the sphere. We show that the hybrid simulation with Botlzmann electrons matches the charging process from Full PIC simulation but starts to deviate at $t = 40 \: \mathrm{\mu s}$ as it was seen in the test case~\autoref{fig:41_plate_graph} for high cell size $\Delta C$. Such effect might be the same for this case and finer mesh might be required for more accurate results. In~\autoref{fig:43_short_graph}b) and c) we have the direct and indirect force estimations. Unfortunately, due to the boundary conditions, the potential field oscillated during the simulation which highly distorted the direct force calculation, but we can see that it manages to follow the same evolution in the indirect force calculation. It is important to note that the simulation has not reached a steady state solution. This case only serves to capture the differences between Full and Hybrid PIC in initial charging. We can conclude that the Boltzmann solver implemented in PANTERA is capable to capture the plasma interaction with satisfactory accuracy. 

\subsubsection{Charging with Kappa fluid}
\begin{figure}[ht]
    \centering
    \includegraphics[width=0.7\columnwidth,trim={0.1cm 0.4cm 0.1cm 0},clip]{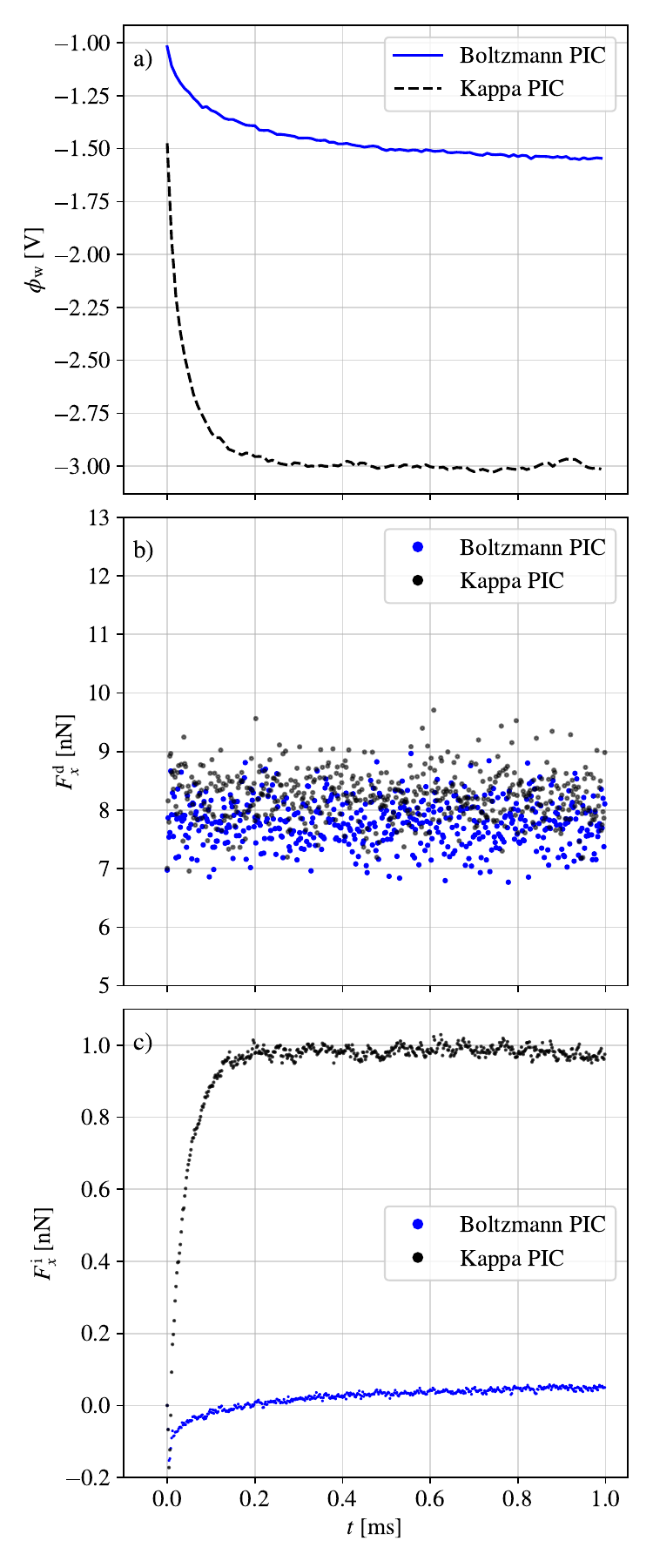}
    \vspace{-4mm}
    \caption{Results from the plasma--sphere interaction simulation for Boltzmann vs Kappa electron fluids. The wall potential in the shadow part of the sphere is shown a) as a function of time. In figure b) we have the direct force and in c) the indirect force.} 
    \label{fig:43_long_graph}
\end{figure}

\begin{figure*}[ht]
    \includegraphics[width=0.75\linewidth]{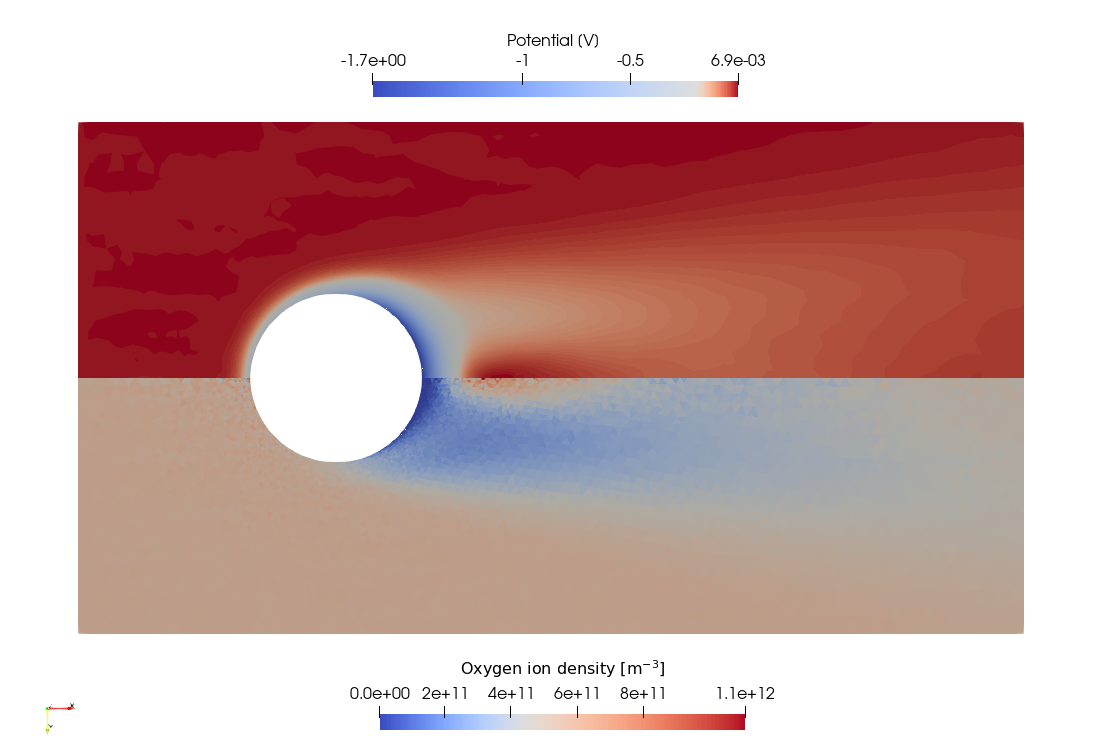} 
    \caption{Snapshot from a simulation of plasma--sphere interaction using Hybrid PIC.}
    \label{fig:43_sphere_diff}
\end{figure*}

Lastly, we show the results for a charging process with Kappa particles to show the effect of the high energy tail. With the non-linear solver equipped in PANTERA which is used for Boltzmann electrons, we attempt to solve the Poisson equations with Kappa electrons. From the Kappa VDF~\eqref{eq:22_kappa_distribution}, one can find the density relation for electrons
\begin{equation}
    n_{\mathrm{e}} (\phi) = n_{0,\mathrm{e}} \left(1 - \frac{e \phi}{k T_{\mathrm{e}} (\kappa-\frac{3}{2})}\right)^{-\kappa + \frac{1}{2}}
\end{equation}
which results in alternative version of the Poisson equation
\begin{equation}
\begin{gathered}
       \varepsilon \int_C \phi_k \grad v_k \grad v_l \dd V + \hfill \\
       e n_{0,j} \int_C  \left(1 - \frac{e \phi_k}{k T_{\mathrm{e}} (\kappa-\frac{3}{2})}\right)^{-\kappa + \frac{1}{2}} v_k v_l \, \dd V  = \\ \sum_{p\in C}  \int_C \delta(\vb{r}-\vb{r}_p) v_l \, \dd V + Q^{\rm acc}_l ,\\
        Q^{\rm acc} = - e \int_{t_0}^t \int_{S}  J^{\mathrm{\kappa}}_{\mathrm{e}} \, \dd S \dd t
\end{gathered}
\end{equation}
where $J^{\mathrm{\kappa}}$ is taken from~\eqref{eq:22_kappa_flux} and $Q^{\rm acc}$ appears only in cells in contact with solid body. It is important to note that only a case of $\alpha=1$ is implemented; no mixture of Boltzmann and Kappa distributions. The results for the comparison between the Boltzmann electrons and Kappa electrons with $\kappa = 4$ are shown in~\autoref{fig:43_long_graph}. At first sight in graph a) we see a significant difference in the wall potential in the shadowed area. Furthermore the wall potential due to the charging of Kappa electrons is as twice as big compared to the potential due to the Boltzmann electrons. This results in the difference of indirect force in graph c), capturing an increase by approximately by ten times, while the direct force remains at the same order. We can estimate following drag coefficients using~\eqref{eq:23_drag_coeff_sep}. For Boltzmann electrons we obtain
\begin{equation}
    C_{\mathrm{c}}^{\mathrm{d}} = 2.34, \quad C_{\mathrm{c}}^{\mathrm{i}} = 0.007
\end{equation}
and for Kappa electrons
\begin{equation}
    C_{\mathrm{c}}^{\mathrm{d}} = 2.49, \quad C_{\mathrm{c}}^{\mathrm{i}} = 0.28
\end{equation}
Thus for different geometries the Kappa distribution are able to charge the surfaces to greater potential which can then result in higher drag and can be comparable to the direct drag. An example from the simulation showing the steady state solution around the sphere is shown~\autoref{fig:43_sphere_diff} with Boltzmann electrons. It is apparent that the sphere due to the flow is not charged uniformly and we obtain greater negative potential at the surface in the shadowed part due to the fact that ions cannot reach these surfaces, unless the electric field would be strong enough to attract them back to the back of the sphere.

\subsection{Satellite flight in ionosphere}
\subsubsection{Satellite geometry and domain setup}
\begin{figure}[ht]
    \centering
    \includegraphics[width=1.\columnwidth,trim={0cm 0cm 0cm 0},clip]{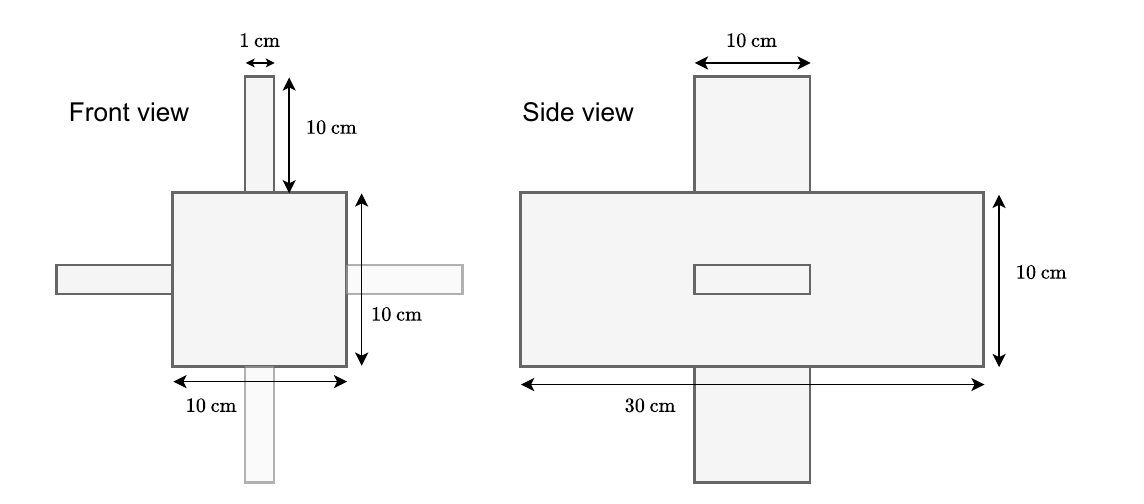}
    \vspace{-4mm}
    \caption{Sketch of a satellite geometry used in the simulation.}
    \label{fig:52_satellite_geometry}
\end{figure}

In~\autoref{fig:52_satellite_geometry} we present the testing geometry of a satellite for the simulation of ionosphere plasma--solid body interaction. The satellite consists of a main body of dimensions $30\,\mathrm{cm} \times 10\,\mathrm{cm} \times 10\,\mathrm{cm}$ with four panels of dimensions $8\,\mathrm{cm} \times 10\,\mathrm{cm} \times 1\,\mathrm{cm}$ at the sides of the satellite. These panels are added to study the effect of thin geometries  for the charged particle collection.

\begin{figure}[ht]
    \centering
    \includegraphics[width=1.\columnwidth,trim={0cm 0cm 0cm 0},clip]{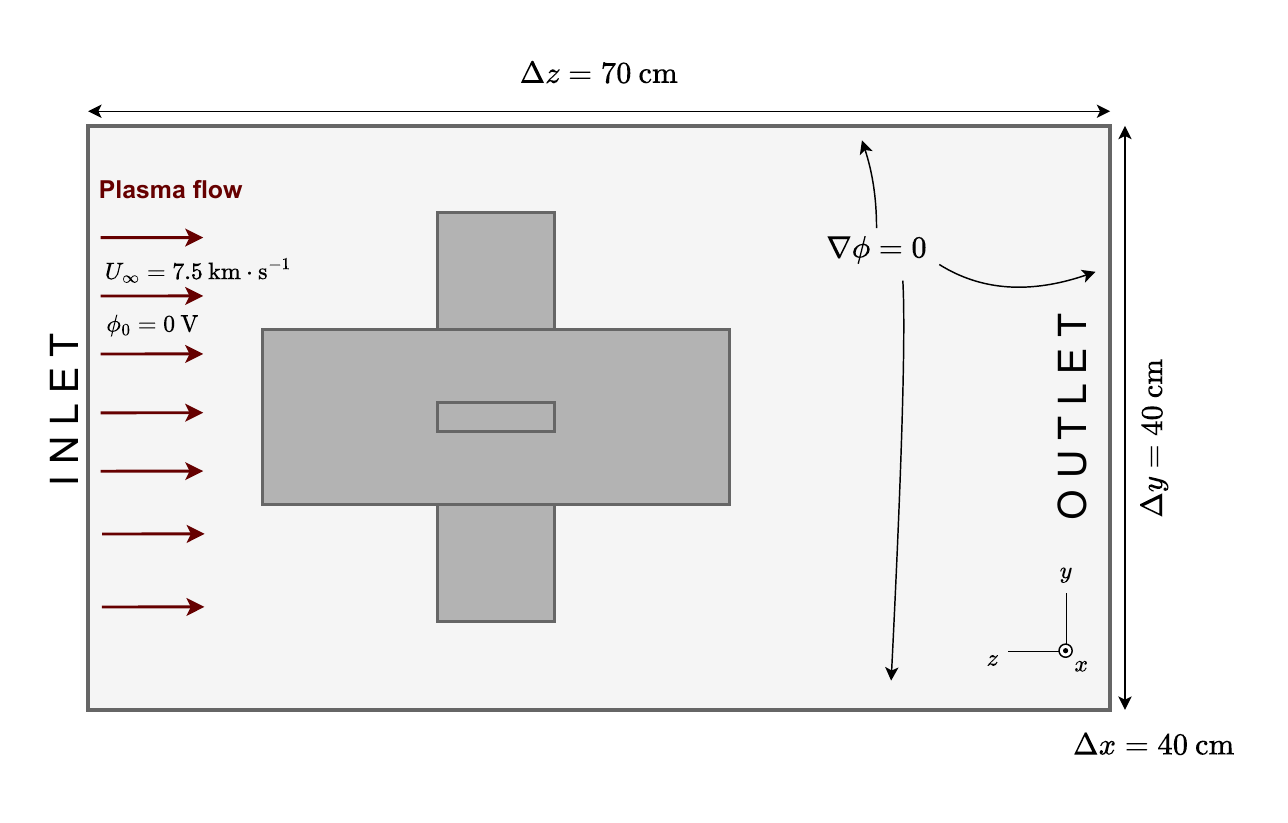}
    \vspace{-4mm}
    \caption{Illustration of a simulation domain for the ionosphere plasma--satellite interaction.}
    \label{fig:52_satellite_domain}
\end{figure}

Such geometry is shown in~\autoref{fig:52_satellite_domain} in a simulation domain consisting of a box $70\,\mathrm{cm} \times 40\,\mathrm{cm} \times 40\,\mathrm{cm}$. We have a relative flow of plasma consisting of oxygen ions O$^+$ and electrons e$^-$ at the inlet with given parameters listed in~\autoref{tab:43_plasma_conditions} representing the orbital conditions at altitude $h = 250$ km. We employ Neumann boundary conditions at the outlet and the sides of the domain. 

\subsubsection{Simulation results}
We now present the obtained simulation results for the case of ionosphere plasma--satellite interaction. In all upcoming results we employed the Hybrid PIC simulation (Boltzmann and Kappa electrons). The mesh in~\autoref{fig:52_satellite_domain} was discretized in similar manner as in the case of the sphere flight. The cell size at the boundaries of the domain starts at $\Delta C = 5 \: \mathrm{cm}$ while it is refined to $\Delta C = 0.2 \: \mathrm{cm}$ in the vicinity of the satellite body to resolve the Debye sheath. The surfaces of the satellite are set to be dielectric to collect all incoming charged particles from the plasma that hit the surface.

\begin{figure}[ht]
    \centering
    \includegraphics[width=1.\columnwidth,trim={0cm 0cm 0cm 0},clip]{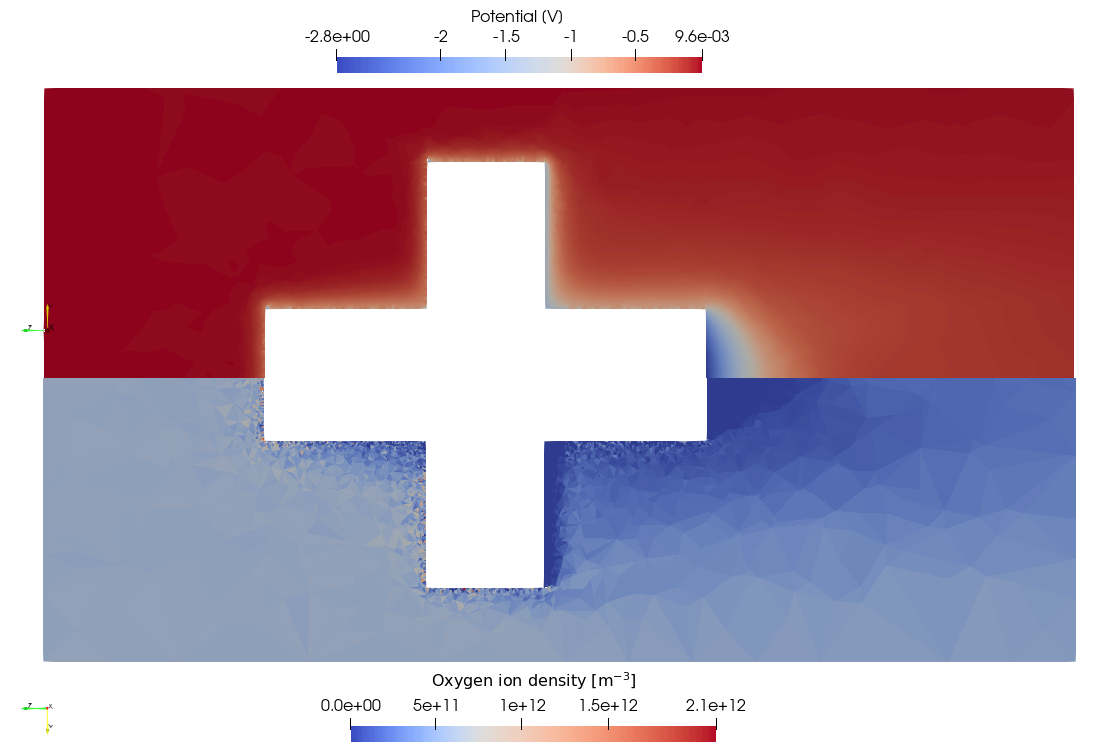}
    \vspace{-4mm}
    \caption{Result from the simulation with Boltzmann electrons on satellite charging. The top half shows potential distribution and the other half shows the ion density.}
    \label{fig:52_boltz}
\end{figure}

The first simulation studied the charging of the satellite using the Hybrid PIC method with Boltzmann electrons and the resulting drag force acting on the body. The result of the potential field and oxygen ion density is presented in~\autoref{fig:52_boltz}. It can be directly seen that the shadowed part (parts covered from the flow of ions) are charged to greater negative potential due to the absence of the ions. The panels, being thin objects, also present the potential difference across the sides. 

\begin{figure}[ht]
    \centering
    \includegraphics[width=0.6\columnwidth,trim={0cm 0cm 0cm 0},clip]{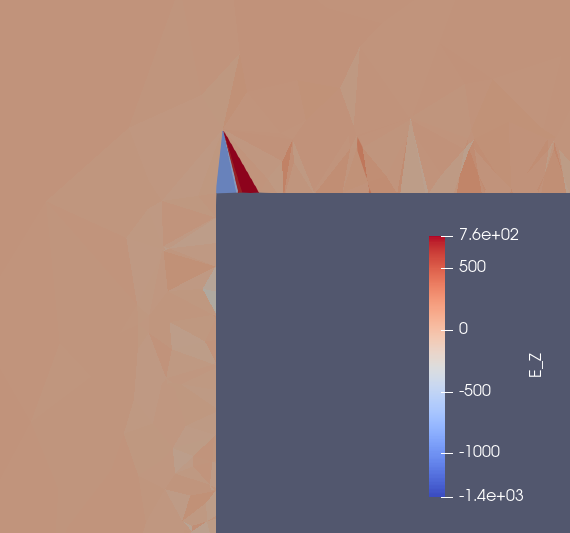}
    \vspace{-4mm}
    \caption{Example of electric intensity peaks at the corners of the satellite geometry.}
    \label{fig:52_peaks}
\end{figure}

A problem arises at the corners of the satellite geometry. A peak in the value of the electric field evolves at the edge of the surface as shown in~\autoref{fig:52_peaks}. The main problem is that such peak can contribute to the calculation of the indirect force according to the Maxwell stress tensor~\eqref{eq:23_maxwell_stress} even though it is a surface integral only through 1-2 cells (but present on all edges of the satellite). It could still have an effect since intensity peak is much greater than all values of the intensity in every cell of the domain. The origin of such peak might be primarly due to the problems of the meshing around sharp edges itself. It is known in the scientific community that such edges produce a problem in the potential distribution. Another reason could be from the very physical point of view. Sharp edges tend to have large electric fields due to the sharp gradients of the electric potential. This of course can not be confirmed at the moment and more tests are required to rule out one of the proposed cases. Nevertheless, we do not find any evidence of numerical instability andwe have shown that the Boltzmann solver offers faster computing while still being capable of capturing the process of space platform charging.

\begin{figure}[ht]
    \centering
    \includegraphics[width=0.7\columnwidth,trim={0cm 0.1cm 0cm 0},clip]{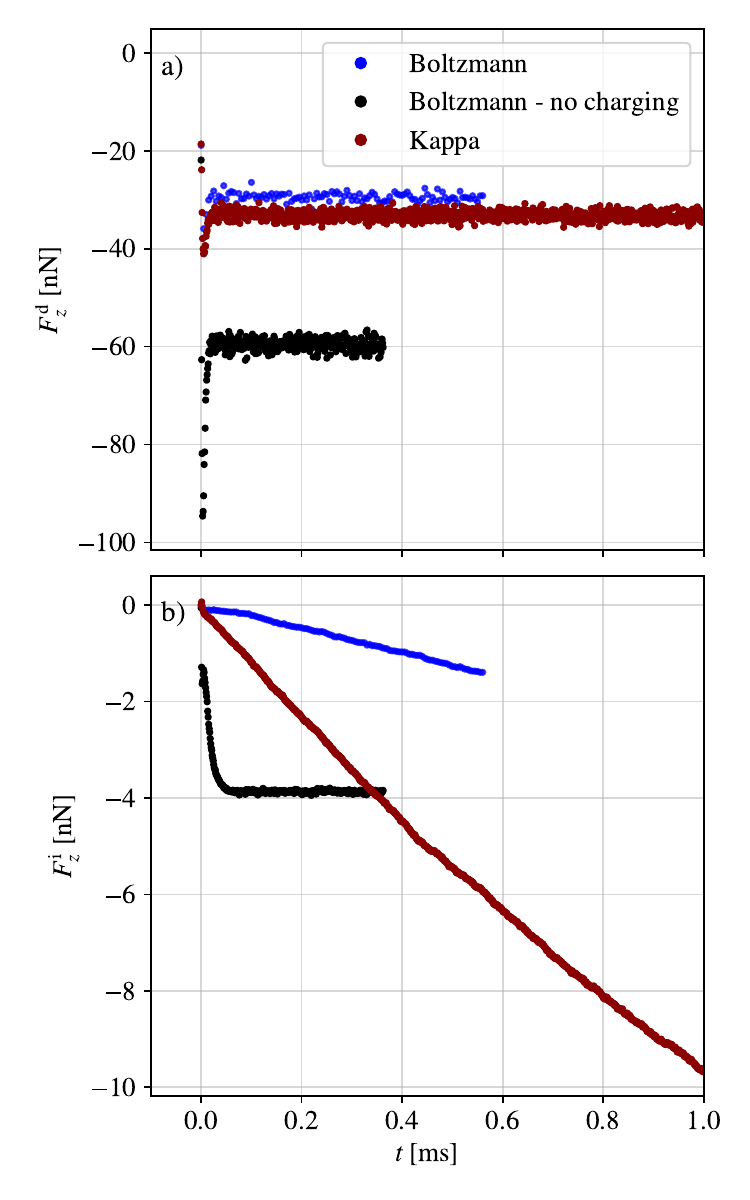}
    \vspace{-4mm}
    \caption{Comparison of a) direct forces and b) indirect forces for all three cases: Boltzmann electrons, Boltzmann electrons with no charging and Kappa electrons.}
    \label{fig:52_forces}
\end{figure}

\begin{figure}[ht]
    \centering
    \includegraphics[width=1.\columnwidth,trim={0cm 0cm 0cm 0},clip]{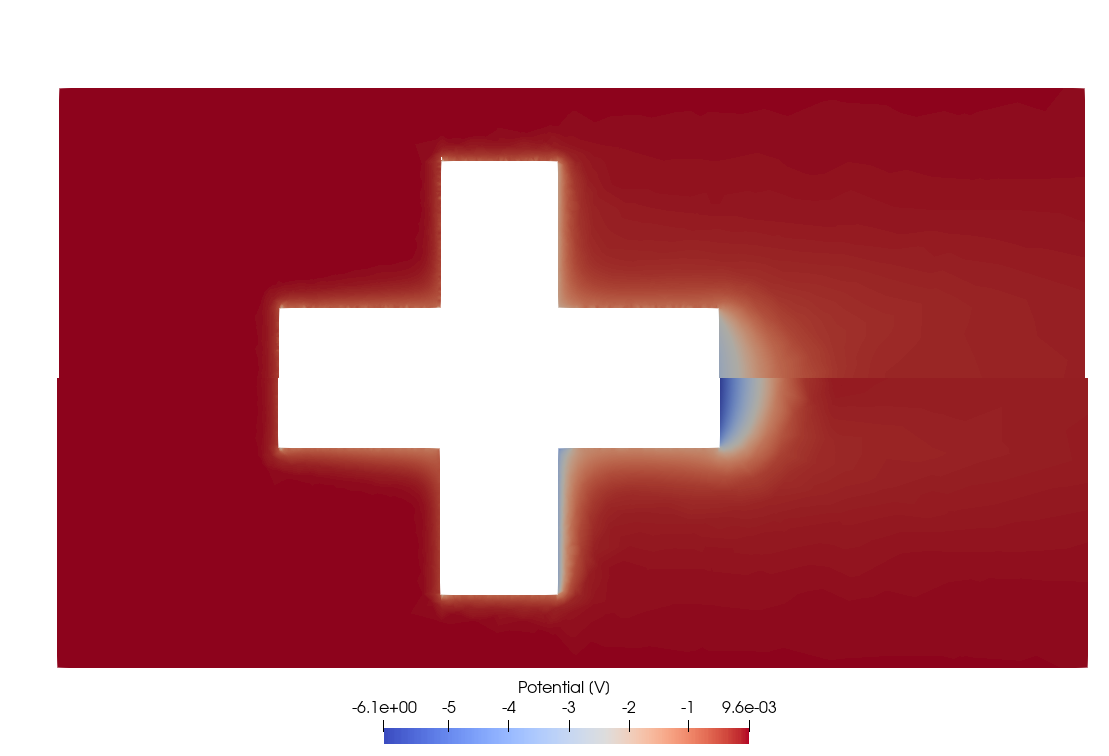}
    \vspace{-4mm}
    \caption{Comparison of the potential field resulting from platform charging between Boltzmann and Kappa electrons.}
    \label{fig:52_kappa}
\end{figure}

Two additional test cases have been simulated. The first one utilized again Boltzmann electrons but the potential on the satellite was kept uniformly constant at value $\phi_{\mathrm{w}} = - 5$ V. In this case, no charging of the platform is simulated. Such case was proposed in order to determine whether the electric intensity peaks are still present. The other case is again with charging processes but with electrons having Kappa distribution to see the differences in overall platform charging. The results from the calcualted forces are shown in~\autoref{fig:52_forces} for all three cases. It can be seen that Boltzmann electrons with no charging have converged to steady state solution while the simulation with charging using both Boltzmann and Kappa distributions are yet in transient regime. The most important result is that we again showed that when electrons with Kappa distribution are present in the ionosphere plasma, the shadowed parts of the satellite get charged to greater negative potential and this further result to increased indirect force, and according to the~\autoref{fig:52_forces}b), the force might be still growing for some time. Thus it is possible that the indirect force might be comparable with direct force, if not greater. In~\autoref{fig:52_kappa} we show the differences in the potential fields between Boltzmann and Kappa electrons developed at the same time step. 

\section{Conclusion \& Future work}\label{sec:6}
In this work we have studied the general mechanism of space platform charging and its relevance to ionospheric aerodynamics of satellites in VLEO orbits. We have implemented a hybrid Fluid-PIC model that removes the time scales of electrons to gain the efficiency in numerical computing. These implemented models were tested on verification cases which were proved to be successful and accurate. Later we simulated space platform charging for two cases: a sphere and a satellite in ionosphere. From both we have obtained interesting results, one of them showing the difference in charging potential when Kappa-dsitributed electrons are considered instead of Maxwell-Boltzmann distributed electrons. It was shown that these electrons cause greater potential differences across the surfaces, thus contributing greatly to the indirect force. While it might seem that the indirect force can become comparable to the direct force, the ionospheric drag still represents a small fraction of the neutral drag that is mainly dominant in the ionosphere.

An improvement to the model implemented in PANTERA is planned to especially account for additional charging mechanisms, such as photoemission, which can affect the process of space platform charging. It is worth to notice that the magnetic field should be accounted for to study its effect on the charging as well. Additionally, an improvement of the satellites geometries is required due to the unwanted electric intensity peaks at the edges that can affect the resulting calculations of drag forces. Lastly, the proposed computing models in PANTERA are planned to be utilized on a real satellite geometry SKIMSAT developed by a space company RedWire to study the effects of the ionosphere on the satellite in great detail.

\vfill
	
	\bibliography{Biblio}

\begin{thebibliography}{35}
\bibitem [{\citenamefont {May}()}]{may}%
  \BibitemOpen
  \bibfield  {author} {\bibinfo {author} {\bibfnamefont {G.}~\bibnamefont {May}},\ }\bibfield  {title} {\bibinfo {title} {Numerical methods for {PDE}},\ }\href@noop {} {\ }\BibitemShut {NoStop}%
\end{thebibliography}



\section*{Appendix  }

\textbf{Appendix A: Derivation of equations for 1D Plasma sheath verification case}

In the work, a verification test case was presented to test the implemented models of Boltzmann solver and indirect force calculation. The proposed case involved additional expansion of equations of the original ones obtained from a literature Bittencourt, 1980.

The equations represent a simplified formation of a plasma sheath at the wall in 1D. One has to start from Poisson equation
\begin{equation}\label{eq:A_poisson}
    \varepsilon \dv[2]{\phi}{x} = e \left( n_{\mathrm{e}(x)} - n_{\mathrm{i}}(x)\right)
\end{equation}
where $\phi$ is the potential field (seeked solution), $\varepsilon$ is the permittivity, assumed to be constant, $e$ is the elementary charge and $n_{\mathrm{e,i}}$ is the electron and ion density, both expressed by the Boltzmann relation
\begin{equation}\label{eq:A_boltz_relation}
        n_{\mathrm{e,i}} = n_0 \exp \left[ \frac{\pm e \phi(x)}{k T_0}\right]
\end{equation}
where $n_0$ is the reference density of quasi-neutral plasma, $k$ is the Boltzmann constant and $T_0$ is the reference temperature (common for both ions and electrons). The latter equation is then substitued back in~\eqref{eq:A_poisson} to obtain
\begin{align}
    \varepsilon \dv[2]{\phi}{x} &= 2 e n_0 \sinh\left(\frac{e\phi}{kT_0}\right)\label{eq:A_differential_equation} \\
    \phi(0) &= \phi_{\mathrm{w}}, \quad \phi(\infty) = 0 
\end{align}
where the BC for infinity represents quasi-neutral plasma. The identity $2 \sinh(x) = \exp(x) + \exp(-x)$ was used. In the referenced book, the original differential equation is obtained by introducing non-dimensional variables
\begin{equation}
    u = \frac{e \phi}{k T_0}, \quad \xi = \sqrt{2} \frac{x}{\lambda_{\mathrm{D}}}, \quad \lambda_{\mathrm{D}} = \sqrt{\frac{\varepsilon k T_0}{n_0 e^2}}
\end{equation}
where $\lambda_{\mathrm{D}}$ is the Debye length. By modifying the constants in~\eqref{eq:A_differential_equation} we get
\begin{equation*}
\frac{1}{2} \frac{\varepsilon k T_0}{n_0 e^2} \dv[2]{}{x} \left( \frac{e\phi}{k T_0}\right) = \sinh\left(\frac{e\phi}{kT_0}\right)
\end{equation*}
which finally gives us
\begin{align}
    \dv[2]{u}{\xi} &= \sinh(u) \label{eq:A_nondifferential_equation} \\
    u(\xi=0) &= u_{\mathrm{w}}, \quad u(\infty) = 0 \label{eq:A_non_bc}\\
\end{align}
where $u_{\mathrm{w}} = e\phi_{\mathrm{w}}/kT_0$. We further continue in this form to obtain the solution. The next step is to multiply~\eqref{eq:A_nondifferential_equation} by $2 \dv{u}{\xi}$ to obtain
\begin{equation}
    2 \dv{u}{\xi} \dv[2]{u}{\xi} = 2 \dv{u}{\xi} \sinh(u)
\end{equation}
and after simple rewritting, one finds that
\begin{equation}
    \dv{}{\xi} \left( \left(\dv{u}{\xi}\right)^2 \right) = \dv{}{\xi} \left( 2 \cosh(u) \right)
\end{equation}
which is then integrated to obtain
\begin{equation}
    \left(\dv{u}{\xi}\right)^2 = 2\cosh(u) + C_0
\end{equation}
by assuming the following BC in~\eqref{eq:A_non_bc} one finds that $C_0 = -2$. We end up with differential equation of the first order
\begin{equation}
    \dv{u}{\xi} = \pm \sqrt{2\left(\cosh(u) -1 \right)}
\end{equation}
where $\pm$ comes from the square root of both sides. Since we wish to deal with negatively charged wall $u_{\mathrm{w}} < 0$, we choose the $+$ sign, which tells us that the potential only increases when going to $\xi = \infty$. We further get
\begin{equation}
    \frac{\dd u}{\sqrt{\cosh(u) - 1}} = \sqrt{2} \dd \xi
\end{equation}
and by integrating with change of variables we come to
\begin{equation}
    \int \frac{\dd u'}{\sqrt{\cosh(u')-1}} = \sqrt{2} (\xi - \xi_0)
\end{equation}
where $\xi_0$ is a constant that will be later found from the boundary condition at the wall. The integral on the left hand side is known and it is shown to be evaluated as
\begin{equation}
    - \sqrt{2} \ln\left(\frac{\exp\left(\frac{u}{2}\right) + 1}{\exp\left(\frac{u}{2}\right) - 1} \right) = \sqrt{2} (\xi - \xi_0)
\end{equation}
which after simple treatment yields the final solution
\begin{equation}
    u(\xi) = -4 \arctanh \left( \exp\left(- (\xi - \xi_0) \right) \right)
\end{equation}
where from the boundary condition $u(0) = u_{\mathrm{w}}$ we find that
\begin{equation}
    \xi_0 = \ln \left( \tanh \left( -\frac{u_{\mathrm{w}}}{4} \right) \right)
\end{equation}
then finally
\begin{equation}
    u(\xi) = -4 \arctanh \left( g(u_{\mathrm{w}}) e^{-u} \right)
\end{equation}
where $g(u_{\mathrm{w}}) = \tanh \left( -\frac{u_{\mathrm{w}}}{4} \right)$ for clearance. The dimensional version follows
\begin{equation}
    \phi(x) = -\frac{4 k T_0}{e} \arctanh\left(g\left(\frac{e\phi_{\mathrm{w}}}{kT_0}\right) e^{-\frac{2}{\lambda_{\mathrm{w}}}} x\right)
\end{equation}The main expansion comes now when we wish to account for the charging of the wall in such way that $u_{\mathrm{w}} = f(t)$ is a function of time. This is achieved by introducing a change in the surface charge given by
\begin{equation}\label{eq:A_surface_charge}
    \dv{\sigma_{\mathrm{w}}}{t} = - \frac{1}{4} e n_0 \underbrace{\sqrt{\frac{8 k T_0}{\pi m_{\mathrm{e}}}}}_{\overline{u}_{\mathrm{e}}} \exp(\frac{e\phi_{\mathrm{w}}}{kT_0})
\end{equation}
where $\sigma_{\mathrm{w}}$ is the surface charge and can be evaluated from the jump condition for the electric field at the wall as follows
\begin{equation}
    \varepsilon E\big |_{x=0} - \varepsilon_{\rm in} E_{\rm in} \big|_{x=0} = \sigma_{\mathrm{w}}
\end{equation}
where $E_{\rm in}$ is the electric intensity inside the wall and is assume to be zero $E_{\rm in} = 0$ and $E = -\dv{\phi}{x}$. This yields the expression for surface charge
\begin{equation}
    \sigma_{\mathrm{w}} = \varepsilon\frac{4 k T_0}{e} \frac{1}{\sqrt{2} \lambda_{\mathrm{D}}} \sinh\left(\frac{e\phi_{\mathrm{w}}}{2kT_0}\right)
\end{equation}
which is put back in~\eqref{eq:A_surface_charge} and derivated with respect to time
\begin{equation}
    \sqrt{2}{\lambda_{\mathrm{D}}} \cosh\left(\frac{e\phi_{\mathrm{w}}}{2kT_0}\right) \dv{\phi_{\mathrm{w}}}{t} = -\frac{1}{4} \frac{e n_0}{\varepsilon} \overline{u_{\mathrm{e}}} e^{\frac{e \phi_{\mathrm{w}}}{k T_0}}
\end{equation}
where for the reminders $\overline{u_{\mathrm{e}}}$ is the thermal speed, not the non-dimensional solution $u$. We now solve this equation using variable separation method, which yields
\begin{equation}
    \int^{\phi_{\mathrm{w}}(t)}_0 \cosh\left(\frac{e\phi'_{\mathrm{w}}}{2kT_0}\right) e^{\frac{e \phi'_{\mathrm{w}}}{k T_0}} \dd \phi'_{\mathrm{w}} = -\frac{\overline{u}_{\mathrm{e}}}{\sqrt{2} \lambda_{\mathrm{D}}} t 
\end{equation}
where the integral is a bit complex to solve but can be solved using tables for integrals. The solution reads
\begin{equation}
    \frac{z^3}{3} + z = D
\end{equation}
where we introduced $z = \exp\left(-\frac{u}{2}\right)$ and $D = \frac{4}{3} + \frac{\overline{u}_{\mathrm{e}}}{\sqrt{2} \lambda_{\mathrm{D}}} t$. This is essentially a cubic equation with unique solution, again using the known solution for cubic equation, one obtains finally
\begin{equation}\label{eq:41_1q}
    \phi_{\mathrm{w}}(t) = - \frac{2kT_0}{e}\ln B(t)
\end{equation}
where
\begin{align}
    B(t) &= \left(\frac{3}{2}D + \sqrt{\frac{9}{4}D^2 + 1}\right)^\frac{1}{3} + \left(\frac{3}{2}D - \sqrt{\frac{9}{4}D^2 + 1}\right)^\frac{1}{3} \label{eq:41_2q}\\
    D &= a_0 + \sqrt{2} \frac{\overline{u}_{\mathrm{e}}}{\lambda_{\mathrm{D}}} t \label{eq:41_3q}\\
    a_0 &= \exp(- \frac{e \phi_{\mathrm{w}}(0)}{2kT_0}) + \frac{1}{3}\exp(- \frac{3 e \phi_{\mathrm{w}}(0)}{2kT_0}) \label{eq:41_4q}
\end{align}
This gives the complete solution to a 1D plasma sheath with charging wall. One can then calculate the indirect force by simply evaluating $E |_{x=0}$ with the wall potentail that is a function of time now. Then using the Maxwell stress tensor
\begin{equation}
    F^{i} = \int_S \frac{1}{2} \varepsilon E^2 \dd S 
\end{equation}
gives the total indirect drag force acting on the wall in 1D.

The obtained data from simulations are easily retrieved from PANTERA in text files that can be later used for postprocessing.

\end{document}